\documentclass[11pt]{article}

\usepackage[width=6.25in, height=9in, centering=true]{geometry}
\usepackage[usenames,dvipsnames]{xcolor}
\usepackage[numbers]{natbib}
\usepackage[unicode=true,bookmarks=false,colorlinks=true,linkcolor=blue,citecolor=blue,urlcolor=blue,breaklinks=true,pdfstartview=FitH]{hyperref}
\usepackage{enumitem}
\setlist[enumerate]{leftmargin=*,label*=\arabic*.}
\usepackage{setspace}
\usepackage[para, ragged, symbol]{footmisc}
\usepackage[page]{appendix}
\usepackage{etoolbox}
\makeatletter
\patchcmd{\@maketitle}
  {\ifx\@empty\@dedicatory}
  {\ifx\@empty\@date \else {\vskip1ex \centering\footnotesize\@date\par\vskip1ex}\fi
   \ifx\@empty\@dedicatory}
  {}{}
\patchcmd{\@adminfootnotes}
  {\ifx\@empty\@date\else \@footnotetext{\@setdate}\fi}
  {}{}{}
\makeatother
\usepackage{fancyhdr}

\usepackage{amsthm,amssymb,amsmath}
\usepackage{sectsty}
\allsectionsfont{\usefont{OT1}{phv}{bc}{n}\selectfont}
\usepackage{titling}
\pretitle{\begin{center}\LARGE \bf}
\posttitle{\par\end{center}\vskip .5em}
\usepackage[sc]{mathpazo}
\usepackage[OT1]{fontenc}
\usepackage{bm}
\usepackage{siunitx,fix-cm}
\usepackage{xfrac}
\usepackage{upgreek}
\usepackage{graphicx}
\graphicspath{{./images/}}
\usepackage{caption,subcaption}
\makeatletter
\renewcommand*\subcaption@label{%
  \caption@withoptargs\subcaption@@label}
\makeatother
\AtBeginEnvironment{tabular}{\doublespacing}
\usepackage{multirow}
\usepackage{array}
\newlength{\colw}
\newcolumntype{W}{>{\centering\let\newline\\\arraybackslash\hspace{0pt}}p{\colw}}

\newcommand{\inv}[1]{{#1}^{-1}}
\newcommand{\tr}{\textrm{tr}}
\renewcommand{\vec}{\textrm{vec}}
\newcommand{\ud}{\, \mathrm{d}}
\newcommand{\abs}[1]{\left \vert #1 \right \vert}
\newcommand{\iid}{\stackrel {\textrm{iid}}{\sim}}
\newcommand{\ind}{\stackrel {\textrm{ind}}{\sim}}
\newcommand{\var}{\textrm{var}}
\newcommand{\cov}{\textrm{cov}}

\newcommand{\rv}[3][1]{#2_{#1},\ldots,#2_{#3}}

\newcommand{\N}{\mathcal N}
\renewcommand{\|}{\,|\,}

\newcommand{\tx}{\textrm}
\newcommand{\wt}{\textrm{wt}\%}
\newcommand{\inner}[1]{\langle#1\rangle}
\newcommand{\E}[1]{\mathbb{E}\left[#1\right]}
\newcommand{\s}{\sigma}
\renewcommand{\a}{\alpha}
\renewcommand{\b}{\beta}
\renewcommand{\t}{{\bm\theta}}
\newcommand{\g}{\gamma}
\renewcommand{\d}{\Delta}
\renewcommand{\l}{{\bm\lambda}}
\newcommand{\LL}{{\bm \Lambda}}
\newcommand{\UU}{{\bm \Upsilon}}
\newcommand{\OO}{{\bm\Omega}}

\newcommand{\mmu}{{\bm\mu}}
\newcommand{\SSigma}{{\bm\Sigma}}
\newcommand{\ap}{{\bm\vartheta}}
\newcommand{\asp}{\gamma}
\newcommand{\lsr}{{\bm\varpi}}
\newcommand{\obs}{{\textrm{obs}}}

\newcommand{\X}{\bm X}
\newcommand{\x}{\bm x}
\newcommand{\Y}{{\bm Y}}
\newcommand{\Z}{{\bm Z}}
\newcommand{\V}{{\bm V}}

\newcommand{\R}{{\bm R}}
\renewcommand{\U}{{\bm U}}
\newcommand{\T}{{\bm T}}
\newcommand{\PPsi}{{\bm \Psi}}
\newcommand{\kbt}{k_BT}
\newcommand{\pvpost}{\tx{p}_\tx{post}}

\begin{document}

\title{Model comparison and assessment for single particle tracking in biological fluids}

\date{November 27, 2015}
\author{Martin Lysy$^{\star}$, Natesh S. Pillai$^\dagger$, David B. Hill$^\ddagger$, M. Gregory Forest$^\ddagger$, \\John Mellnik$^\ddagger$, Paula Vasquez$^\S$, Scott A. McKinley$^\P$}
\maketitle

\begin{abstract}
\noindent State-of-the-art techniques in passive particle-tracking microscopy provide high-resolution path trajectories of diverse foreign particles in biological fluids. For particles on the order of \SI{1}{\micro\metre} diameter, these paths are generally inconsistent with simple Brownian motion.  Yet, despite an abundance of data confirming these findings and their wide-ranging scientific implications, stochastic modeling of the complex particle motion has received comparatively little attention.  Even among posited models, there is virtually no literature on likelihood-based inference, model comparisons, and other quantitative assessments.  In this article, we develop a rigorous and computationally efficient Bayesian methodology to address this gap. We analyze two of the most prevalent candidate models for \SI{30} second paths of \SI{1}{\micro\metre} diameter tracer particles in human lung mucus: fractional Brownian motion (fBM) and a Generalized Langevin Equation (GLE) consistent with viscoelastic theory.  Our model comparisons distinctly favor GLE over fBM, with the former describing the data remarkably well up to the timescales for which we have reliable information.\let\thefootnote\relax\footnote{$^\star$University of Waterloo, $^\dagger$Harvard University, $^\ddagger$University of North Carolina at Chapel Hill, $^\S$University of South Carolina, $^\P$Tulane University.}
\end{abstract}


\newpage

\section{Introduction}\label{sec:intro}

Over the last two decades, advances in microscopy have provided unprecedented observations of the fluctuating dynamics of microparticles in biological fluids. An important and ubiquitous finding from many experiments is that microparticle diffusion in biological fluids is not well-described by simple Brownian motion.  The most prominent gauge for the departure from the Brownian regime is the mean squared displacement (MSD) of a particle's trajectory $X(t)$:
\[
\inner{X^2(t)}= \E{\big(X(t) - X(0)\big)^2}, \quad t \ge 0.
\]
While the MSD of ``ordinary'' viscous diffusion scales linearly with time, $\inner{X^2(t)} \propto t$, there is now a preponderance of biological examples of sublinear MSD growth,
\begin{equation}\label{eq:anomsubdiff}
\inner{X^2(t)}\sim t^\alpha,
\end{equation}
for $0 < \alpha < 1$ and $t$ ranging over some experimental time frame. This MSD scaling behavior is known as \emph{subdiffusion}. Examples of this behavior include: Adeno-associated virus in cytoplasm \citep{Seisenberger:2001bv}; lipid granules in living fibroblasts \citep{caspi-granek00} and living yeast cells \citep{TolicNorrelykke:2004bf,jeon-et-al11,verdaasdonk-et-al13};  
tracer particles in F-actin networks \citep{Wong:2004by}; RNA in E.\ coli \citep{golding-cox06}; telomeres in the nucleus of mammalian cells \citep{bronstein-et-al09}; tracers in a Dextran solution intended to mimic diffusion in a crowded environment \citep{ernst-et-al12}; and passive micron diameter beads in human bronchial epithelial cell culture mucus \citep{hill-et-al14}.

Subdiffusion can be due to several physical mechanisms: the frequency-dependent viscous and elastic moduli of a complex fluid medium \citep{mason-weitz95}; hindered motion in a crowded environment~\citep{ernst-et-al12}; long-term trapping events that result from binding with a physically constrained substrate in the fluid~\citep{SAXTON:2007vi}; and caging events that occur due to particle movement within and between pores in an immersed mesh structure~\citep{Wong:2004by}. Therefore, careful stochastic modeling and statistical inference can have an important scientific impact -- not only for investigating the basic mechanism of microparticle movement, but also for its application to diagnosis of disease~\citep{Georgiades:2014ey}, monitoring of disease progression~\citep{hill-et-al14}, and engineering of drug delivery vehicles~\citep{lai-et-al07,Cone:2009fr,ensign2012mucus,schuster-et-al15}.

Mucus is a particularly important and challenging biological fluid to understand.  Thin mucosal layers line the eyes, nasal cavity, lung airways, gut, and female reproductive tract, and stand as the body's first line of defense against foreign toxins, particulates, and pathogens. However, our understanding of how these foreign bodies interact with and move through mucus remains far from complete. The microstructure of human lung mucus, our motivating biological fluid, is strikingly complex. By weight, mucus is mostly water, (\numrange[range-phrase=--]{92}{98} \wt). The remaining portion consists of diverse small molecules including proteins, immunoglobulins, salts, and contents of dead cells (e.g., DNA), along with a family of huge polymeric molecular species called \emph{mucins}~\citep{carlstedt1982isolation,carlstedt1983macromolecular,carlstedt1983isolation,carlstedt1984macromolecular,Thornton:2004jh}. These species collectively assemble in the water solvent to create a heterogeneous three-dimensional mesh, with repulsive or attractive interactions specific to the mucus microstructure, and to the size and surface chemistry of the observed particle.  The theoretical challenge of rigorous, physical modeling of microparticle dynamics in such complex molecular systems is daunting, and not likely to be overcome soon.  Nevertheless, strong signals in particle tracking data have been repeatedly observed.  For example, small entities like antibodies (\SI{5}{\nano\metre} effective radius) and virions (up to \SI{60}{\nano\metre}) have weak interaction with the mucin network, and are well-described by simple Brownian motion~\citep{Saltzman:1994vv,Olmsted:2001wi,chen-et-al14}.   On the other hand, larger amine-modified and carboxylated particles (\SI{200}{\nano\metre} to \SI{5}{\micro\metre} radius) exhibit distinctly subdiffusive behavior~\citep{dawson-et-al03,Suk:2007tw,lai-et-al07,hill-et-al14}.

\subsection{The Problem}\label{sec:problem}

While a sublinear MSD is a clear indicator of non-Brownian movement, the MSD is a summary statistic, and does not preserve other important information contained in the data.  To be specific, biological fluids such as mucus are considerably heterogeneous, due to their diverse molecular composition.  Particle-tracking microrheology provides unprecedented information about this heterogeneity by analyzing particle paths in different spatial locations within the same fluid~\citep{Valentine:2001fx,mellnik-et-al14}.  To this end, several groups have adopted a path-by-path estimate of the MSD that is calculated from discrete observations $\rv [0] X N$ of $X(t)$ at equal time intervals of length $\d t$.  This pathwise MSD estimate is defined as~\citep[e.g.,][]{he-et-al08,lubelski-et-al08,didier-et-al12}
\begin{equation}\label{eq:defn-pathwise-msd}
\widehat{\tx{MSD}}(k \cdot \d t) = \frac 1 {N-k+1} \sum_{n=0}^{N-k} (X_{n+k}-X_n)^2.
\end{equation}
However, even within a homogeneous medium, the MSD alone does not suffice to characterize the dynamics of $X(t)$.  That is it to say, while it might be sufficient \emph{within} a parametric family of statistical models, it is not sufficient \emph{between} models, and this often leads to invalid model selection procedures if one fails to account for ancillary information~\citep{robert-et-al11}.

Many modeling efforts in the recent particle-tracking literature ignore this subtle point, focusing exclusively on matching the MSD and other summary statistics to the observed data rather than modeling the complete stochastic process that produces $X(t)$.  Unfortunately, such an approach has important scientific and biomedical limitations.  Indeed, a primary objective in mucus biology is to predict first-passage times of foreign microparticles through protective mucosal layers. If a pathogen's passage times are shorter/longer than the time for clearance of the mucus barrier, then those pathogens do/do not pose a threat to the underlying tissue. However, it can be shown that two popular models in particle tracking -- fractional Brownian Motion (fBM)~\citep{mandelbrot-vanness68} and the Continuous-Time Random Walk (CTRW)~\citep{metzler-klafter00} -- both exhibit uniform subdiffusion, $\inner{X^2(t)} \propto t^\a$, but have respectively \emph{finite} and \emph{infinite} mean escape times from a bounded interval~\citep{Rangarajan:2000eq,OMalley:2011hy}.  Thus, first-passage time estimates depend critically on model features which are not captured by the MSD alone.

\subsection{Our Contribution}\label{sec:contrib}

While microparticle displacement data continue to proliferate, the development of statistical methods to compare and evaluate different models for these data is an open area of research.  
Our principal aim in this article is to perform rigorous likelihood-based comparisons between two key models of sub\-diffusion: fBM and a Generalized Langevin Equation (GLE) with tunable subdiffusive range described in Section~\ref{sec:models}.   Originally formulated in one dimension, both of these models are adapted here to two-dimensional particle trajectories, with inference strategies specifically designed to control the computational burden (Section~\ref{sec:modelfit}).

\begin{figure}[htb!]
	\centering
	\begin{subfigure}{\textwidth}
		\includegraphics[width=1.00\textwidth]{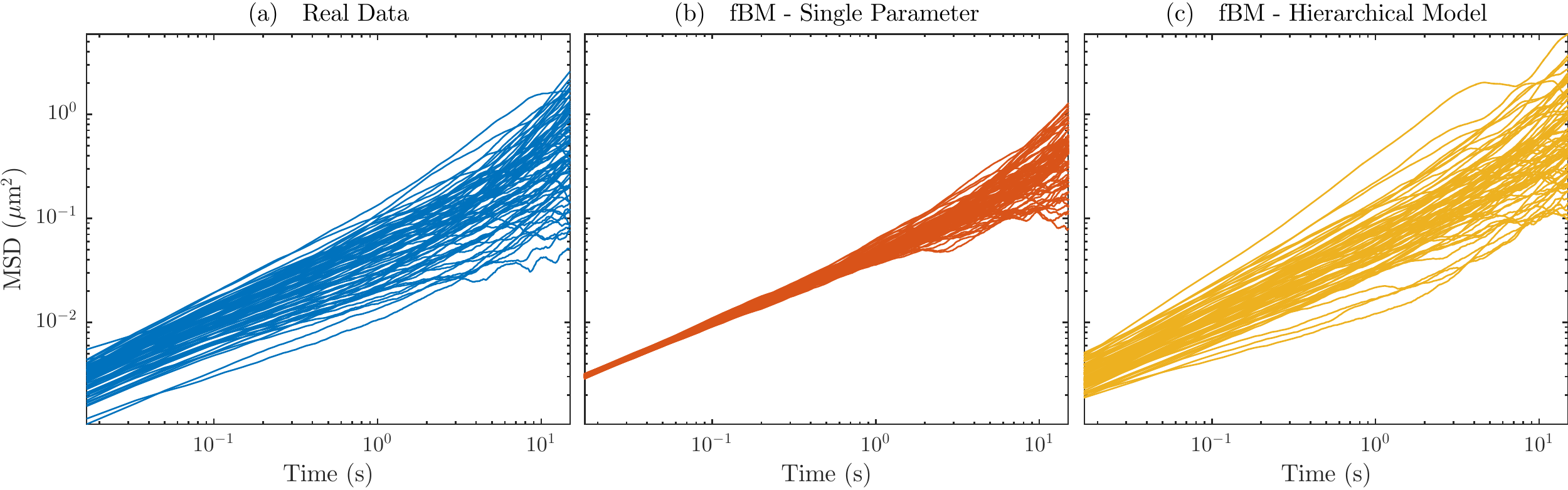}\phantomsubcaption\label{fig:heterdataa}\phantomsubcaption\label{fig:heterdatab}\phantomsubcaption\label{fig:heterdatac}
	\end{subfigure}
	\caption{(a) Ensemble of pathwise MSD estimates for 76 tracer bead paths in 2.5 \wt{} human bronchial epithelial culture mucus.  (b) Pathwise MSDs for 76 simulated paths from a ``homogeneous'' fBM model.  That is, all paths are simulated from a common parameter value, that of the MLE fitted to the 76 trajectories in (a). (c) Pathwise MSDs for simulated paths from a hierarchical fBM model, with each path having its own set of parameters.  The distribution from which these parameters are drawn is described in Section~\ref{sec:modelcomp}.}
	\label{fig:heterdata}
\end{figure}

The data we examine is an ensemble of 76 trajectories of \SI{1}{\micro\metre} diameter polystyrene ``tracer particles'' in 2.5 \wt{} mucus~\citep{hill-et-al14}.  This collection of paths exemplifies the essential problem of particle heterogeneity, as illustrated in Figure~\ref{fig:heterdata}.  
In this data set, the physical environment is controlled to be as homogeneous as is feasible for a mucus sample, with essentially identical immersed microparticles. Despite these controls, the distribution of pathwise MSD estimates (Figure~\ref{fig:heterdataa}) is considerably more widely spread than one would expect from an independent and identically distributed (iid) sample from a uniform population (Figure~\ref{fig:heterdatab}).  This spread in pathwise MSDs has been widely observed for particles in biological fluids and is often referred to as \emph{ergodicity breaking}~\citep{lubelski-et-al08}. This terminology owes to the fact that time-averaged statistics within individual paths do not converge to those of the ensemble average.

In Section~\ref{sec:modelcomp}, we perform model comparisons which harness the full power of the likelihood by computing Bayes factors.  A common criticism of this approach is its sensitivity to the choice of prior distributions, which can influence considerably the preference for one model over the other. 
To address this issue, we utilize a data-driven prior obtained by embedding our collection of sample paths within a hierarchical model. We present an efficient method for approximately fitting such models by conducting the bulk of the computations in parallel.  Our calculations indicate that the GLE model is a better fit, owing to distinct evidence of non-uniform subdiffusion over the experimental timeframe.

Our secondary aim in this paper is to assess which features of the data our posited models capture and which ones they do not.  This is done with a variety of Bayesian predictive checks (Section~\ref{sec:gof}).  In particular, we examine a set of Bayesian model residuals, which are both easily calculable and extremely sensitive to model miss-specification.  We find a remarkable agreement between theoretical and empirical MSDs at short timescales (\numrange[range-phrase=--]{2}{5} s), after which the data provide significantly less reliable information. 

An important tracer particle model we have not analyzed here is the Continuous-Time Random Walk (CTRW). CTRW models attempt to model particle confinement via normal diffusion interspersed with periods of immobilization. When the distribution of the immobilization times is heavy-tailed, CTRW models produce subdiffusive behavior.  
In terms of empirical evidence, the principal argument in favor of CTRW is that it exhibits ergodicity breaking, whereas GLE and fBM trajectories generated from a \emph{single} parameter set do not~\citep{lubelski-et-al08,metzler-et-al09,jeon-metzler10b,jeon-et-al11,burov-et-al11,meroz-et-al13}.  However, a hierarchical fBM or GLE model with a \emph{distribution} of parameters induces ergodicity breaking by definition, and can effectively capture the between-path heterogeneity displayed by our data 
(Figure~\ref{fig:heterdatac}).  We conclude with a discussion of these results and directions for future work in Section~\ref{sec:disc}. 


\section{Overview of Candidate Models} \label{sec:models}
Fractional Brownian motion (fBM) and the Generalized Langevin Equation (GLE) are stochastic, continuous-time models that both have been independently posited and applied to the analysis of subdiffusive behavior.  Both models feature stationary Gaussian increments and continuous sample paths, with some additional properties given below.

\subsection{Fractional Brownian Motion}

A simple model for subdiffusion in the particle displacement process $X(t)$ is given by
\begin{equation}\label{eq:fbm}
X(t) = B_H(t),
\end{equation}
where the fBM process $B_H(t)$ is a zero-mean Gaussian process with covariance
\[
\cov\big(B_H(t), B_H(s)\big) = \tfrac 1 2 \left(\abs{t}^{2H} + \abs{s}^{2H} - \abs{t-s}^{2H}\right), \quad 0 < H < 1.
\]
The Hurst parameter $H$ is used to describe fBM's long-range dependence.  That is, let $x_n = X(n \d t + \d t) - X(n \d t)$ be particle trajectory increments of length $\d t$.  Then
\begin{equation}\label{eq:fbmacf}
\cov(x_n, x_{n+k}) = \tfrac {1} 2 (\d t)^{2H}\left(|k+1|^{2H} + |k-1|^{2H} - 2|k|^{2H}\right).
\end{equation}
While the $x_n$ are not independent,~\eqref{eq:fbmacf} shows that they are stationary.  For $H = 1/2$, fBM reduces to standard Brownian motion.  For $H \neq 1/2$, fBM has long-range memory effects, in the sense that~\eqref{eq:fbmacf} has power law decay.  It is noteworthy that fBM is the unique stochastic process having (i) continuous paths, (ii) stationary (but dependent) increments, and (iii) possessing the self-similarity property: $B_H(at) \stackrel{D}{=} \abs a^H B_H(t)$.  Moreover, fBM is the only Gaussian process with (i) and (ii) exhibiting \emph{uniform subdiffusion}: $\inner {X^2(t)} = t^{\a}$ for all $t > 0$, with subdiffusion coefficient $\a = 2H < 1$.


\subsection{A Generalized Langevin Equation for Viscoelastic Subdiffusion}

Our second candidate model for a subdiffusive tracer particle is constructed directly from principles of statistical mechanics and viscoelastic theory.  This model for the trajectory $X(t)$ admits the explicit path representation~\citep{mckinley-et-al09}
\begin{equation}\label{eq:gle0}
X(t) = C_0 B_0(t) + \sum_{j=1}^{K-1} C_j Y_j(t),
\end{equation}
where $B_0(t)$ is Brownian motion, and $\ud Y_j(t) = -r_j Y_j(t) \ud t + \ud B_j(t)$ are independent Ornstein-Uhlenbeck (OU) processes independent of $B_0(t)$.  The stationary Gaussian increments of $X(t)$ have autocorrelation
\begin{equation}\label{eq:gleacf}
\cov(x_n, x_{n+k}) = C_0^2 \d t \delta_k - \sum_{j=1}^{K-1} C_j^2 \big(e^{-r_j\d t \abs{k+1}} + e^{-r_j\d t \abs{k-1}} - 2e^{-r_j \d t \abs k}\big).
\end{equation}
Expressions for $C_0$, $C_j$, and $r_j$ are provided in the Supplementary Material (\ref{sec:glecov}), as functions of two parameters $\asp$ and $\tau$. Together with $K$, these produce an MSD with tunable subdiffusive range as we describe momentarily.

The Generalized Langevin Equation (GLE) was originally proposed as a model for viscoelastic dynamics by~\cite{mason-weitz95}, founded on the classic work by~\cite{kubo66}.  A detailed account of the GLE aimed for a statistical audience is given by~\cite{kou08}.  A GLE takes the form of a stochastic integro-differential equation for the particle velocity $V(t) = \dot X(t)$:
\begin{equation}\label{eq:glem}
\dot V(t) = - \int_{-\infty}^t \phi(t-s) V(s) \ud s + F(t),
\end{equation}
where $\dot V(t)$ is the particle acceleration, and $F(t)$ is a stationary mean-zero Gaussian process with $\cov\big(F(t), F(s)\big) \propto \phi(\abs{t-s})$ (see~\ref{sec:glecov} for details). Several authors have discussed formulations of the ``memory kernel'' $\phi(t)$ for which the GLE exhibits subdiffusion~\citep{morgado-et-al02, kou-xie04, kneller11}.  Here, we follow~\cite{mckinley-et-al09} and employ the \emph{generalized Rouse kernel}
\begin{equation}\label{eq:rouse-kernel}
\phi(t) = \phi_K(t; \asp, \tau) = \frac{1}{K}\sum_{k=1}^K \exp(-\abs{t}/\tau_k), \qquad \tau_k = \tau \cdot (K/k)^\asp.
\end{equation}
The decomposition of $\phi(t)$ into a sum of exponentials is a long-standing model in linear viscoelastic theory~\citep[e.g.,][]{soussou-et-al70, ferry1980viscoelastic}.  
Furthermore, for particles like ours with negligible mass, such a decomposition provides the otherwise intractable GLE with the explicit solution~\eqref{eq:gle0}.

Each $\tau_k$ in~\eqref{eq:rouse-kernel} corresponds to a distinct ``relaxation time'' of the viscoelastic system.  However, unconstrained relaxation times for all but very small $K$ cannot be estimated reliably from the typical amount of data at hand~\citep{fricks-et-al09}.  Moreover, for larger $K$ it is natural to think of the $\tau_k$ as approximating a \emph{continuous} relaxation spectrum, which can be parametrized parsimoniously. 
Indeed,~\cite{mckinley-et-al09} have shown that the GLE with Rouse kernel~\eqref{eq:rouse-kernel} exhibits \emph{transient subdiffusion}:
\begin{equation}\label{eq:trans-subdiff}
\inner{X^2(t)} \sim \begin{cases} t^{\a} & t_0 < t < t_1, \\
t & t > t_1. \end{cases}
\end{equation}
The subdiffusion coefficient is given by $\a = 1/\asp$.  The process ``transitions'' to ordinary diffusion for $t > t_1$, with $t_1 \to \infty$ as $K$ increases.  
The parameter $\tau$ is the timescale of ``shortest memory'': the smallest timescale at which the particle interacts with its environment.  It also prescribes a time-scaling law for the MSD via $\inner{X^2(t) \| \asp, \tau} = \inner{X^2(t/\tau) \| \asp, 1}$.  Thus, the subdiffusive range $(t_0, t_1)$ is implicitly determined by $K$ and $\tau$.



\section{Data Collection and Model Fitting}\label{sec:modelfit}

\subsection{Data Collection}\label{sec:data}

Our data collection methods are fully detailed in~\cite{hill-et-al14}.  To summarize, \SI{1}{\micro\metre} polystyrene particles with carboxyl surface chemistry were placed in \SI{1}{\centi\metre} discs, each containing \SI{5}{\micro\liter} of mucus 
harvested from primary human bronchial epithelial (HBE) cell cultures. 
The bead surface treatment was chosen to minimize binding and repulsion affinities between the beads and mucins in the fluid environment. The motion of the beads was recorded at 60 frames per second for 30 seconds each.  Subsequently, the bead position within each camera frame was determined by the \emph{Video Spot Tracker} software (\url{http://cismm.cs.unc.edu/downloads/}).

\begin{figure}[htb!]
	\centering
		\includegraphics[width=1.00\textwidth]{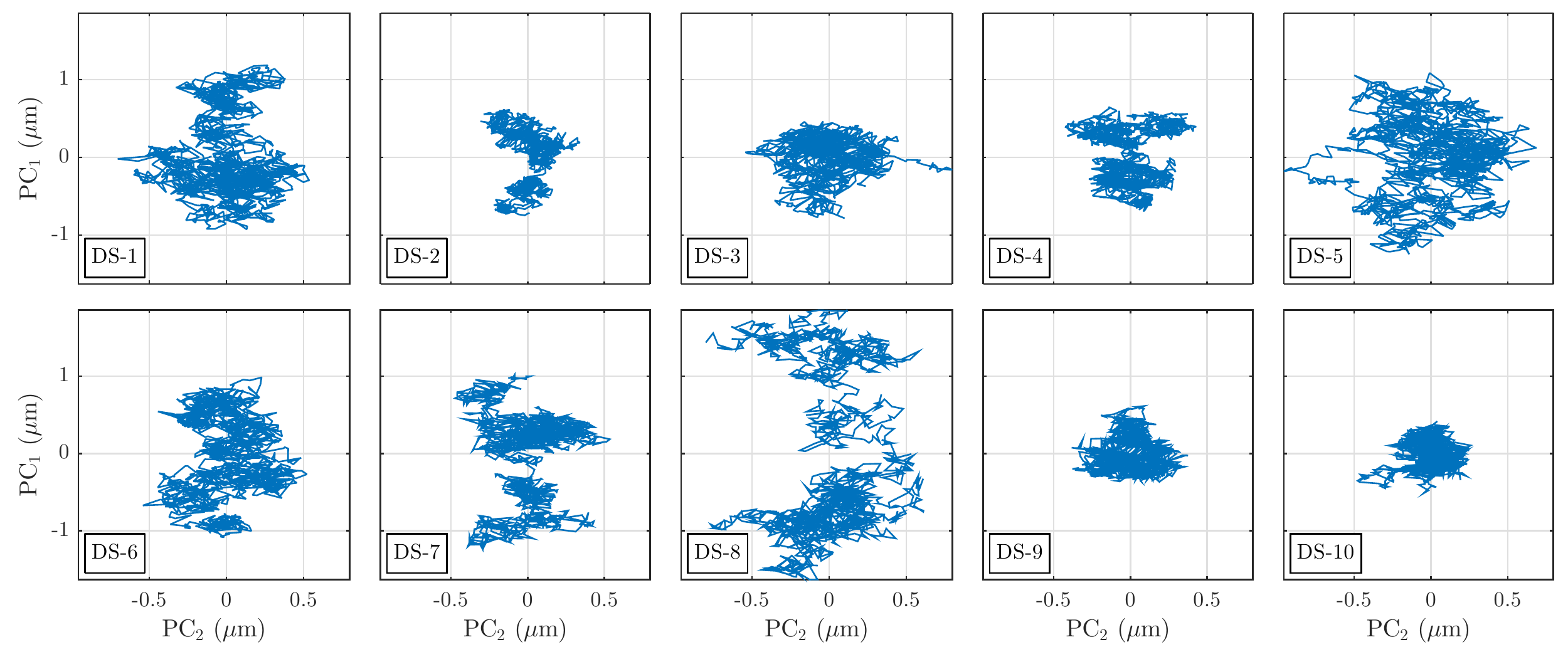}\\
		\includegraphics[width=1.00\textwidth]{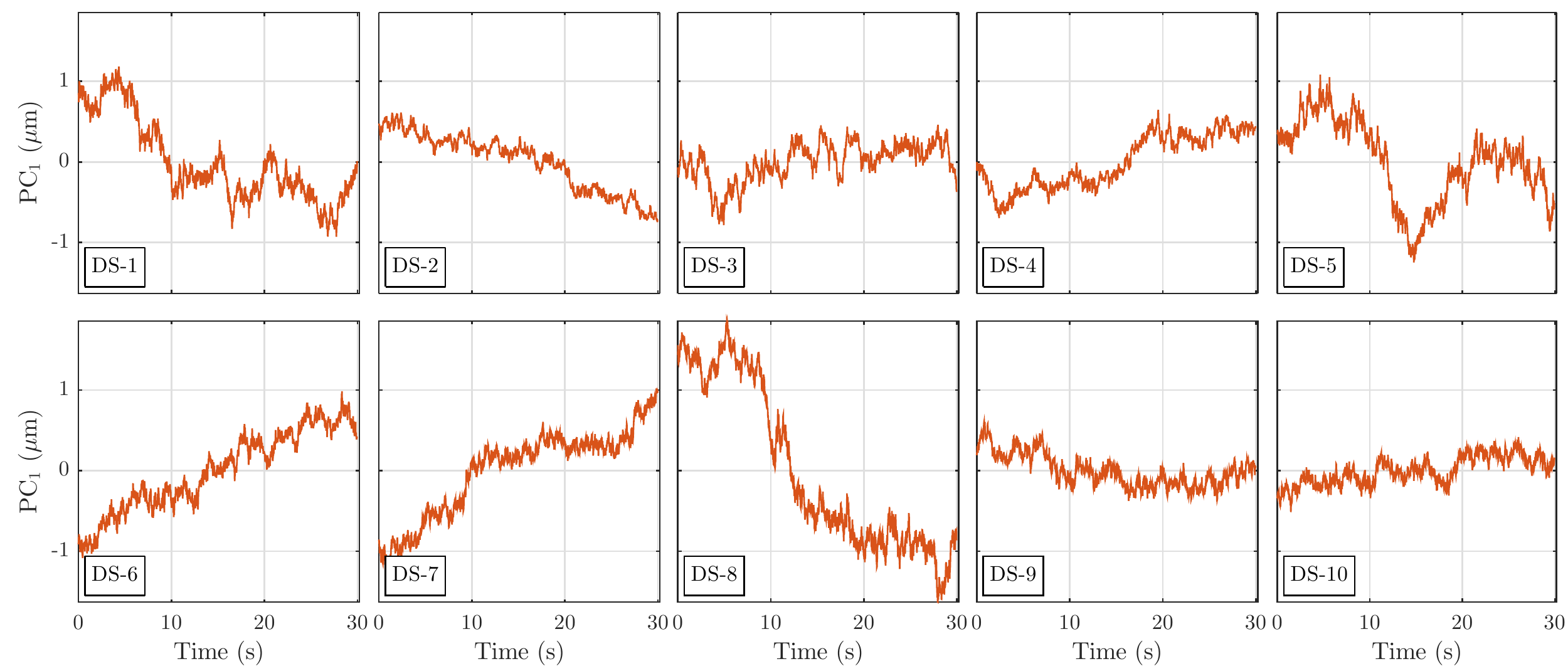}
	\caption{Displacements curves $\X(t)$ for 10 representative bead trajectories with 2.5 \wt{} mucus concentration.  The trajectories have been shifted to have mean zero and rotated such that $\tx{PC}_1$ and $\tx{PC}_2$ are the orthogonal directions of greatest and least movement (principal components).  The top two rows display two-dimensional trajectories, while the bottom rows display the movement along $\tx{PC}_1$ as a function of time.}
	\label{fig:data}
\end{figure}

Experimental particle trajectories were obtained for various levels of mucus concentration, ranging from 1 \wt{} to 5 \wt.  Motion in 1 \wt{} mucus (very dilute) was similar to Brownian motion, while particle paths in 5 \wt{} mucus exhibited a wide range of behaviors.  The most uniform group showing persistent non-Brownian behavior was for 2.5 \wt{} mucus, and so we used this group for the analysis which follows.  A total of 76 two-dimensional trajectories in 2.5 \wt{} mucus were recorded, ten of which are displayed in Figure~\ref{fig:data}.

\subsection{Model Fitting}\label{sec:fit}

Let $\X(t) = (X_1(t), X_2(t))$ denote the particle's two-dimensional position at time $t$, and $\X = (\rv [0] {\X} N$) denote the recorded observations with $\X_n = \X(n\d t)$, $\d t = 1/60$, and $N = 1800$.  We embed our candidate subdiffusion models in a location-scale model of the form
\begin{equation}\label{eq:lsmod}
\X(t) = \mmu t + \SSigma^{1/2} \bm{Z}(t),
\end{equation}
where $\mmu = (\mu_1, \mu_2)$ accounts for linear drift, $\SSigma = \left[\begin{smallmatrix}\sigma_{1}^2&\sigma_1\sigma_2\rho \\ \sigma_1\sigma_2\rho &\sigma_2^2 \end{smallmatrix}\right]$ is a variance matrix, and $\bm{Z}(t) = (Z_1(t), Z_2(t))$ are iid copies of either the fBM model~\eqref{eq:fbm} or the GLE model~\eqref{eq:gle0}.  The measurement error in our observations was found to be negligible (\ref{sec:error}), and therefore we assume that the recorded data are actual particle positions.

Both the fBM and GLE models are nonstationary but have stationary increments. That is, $\x_{N\times 2} = (\rv {\x} N)$ with $\x_n = \X_n-\X_{n-1}$ are observations of a discrete-time stationary Gaussian process.  The increments have a matrix-normal distribution
\begin{equation}\label{eq:lsincr}
\x \sim \N_{N\times 2} \left({\bm \d t} \mmu, {\bm V}_{\ap}, \SSigma\right), 
\end{equation}
for which the log-likelihood is
\begin{equation}\label{eq:llmod}
\ell(\mmu, \SSigma, \ap \| \x) = -\frac 1 2 \left\{\tx{tr}\left[\bm{V}^{-1}_{\ap}(\x - \bm{\d t}\mmu)' \SSigma^{-1}(\x - \bm{\d t}\mmu )\right] + N \log \abs{\SSigma} + 2 \log \abs{\bm{V}_{\ap}} \right\},
\end{equation}
where $\bm{\d t}_{N\times 1} = (\d t, \ldots, \d t)$, and $\ap$ are the parameters of the subdiffusion models.  Specifically, $\bm{V}_{\ap}$ is the Toeplitz variance matrix corresponding to the fBM autocorrelation~\eqref{eq:fbmacf}, or the GLE autocorrelation~\eqref{eq:gleacf}, such that $\ap_\tx{fBM} = H$ and $\ap_\tx{GLE} = (\asp, \tau)$.  Working with the stationary increments $\x$ instead of positions $\X$ reduces the cost of each log-likelihood evaluation from $\mathcal O(N^3)$ to $\mathcal O(N^2)$ via the Durbin-Levinson algorithm~\citep[e.g.,][]{brockwell-davis09}.

To estimate the parameters of~\eqref{eq:lsmod} in a Bayesian setting, we employ the prior distributions
\begin{align}\label{eq:specprior}
\pi_\tx{fBM}(H) \propto 1 && &\pi_\tx{GLE}(1/\asp) \propto 1 \nonumber\\
\pi_\tx{fBM}(\mmu, \SSigma \| H) \propto \abs{\SSigma}^{3/2} &,& &\log(\tau) \| \asp \sim \N(-6.91, 2.68^2) \\
&& & \pi_\tx{GLE}(\mmu, \SSigma \| \asp, \nu) \propto \abs{\SSigma}^{3/2}.\nonumber
\end{align}
The noninformative priors on $(\mmu, \SSigma)$ are independence-Jeffreys priors~\citep{sun-berger07}, and a Lebesgue prior is given to the subdiffusion parameters $\a_\tx{fBM} = 2H$ and $\a_\tx{GLE} = 1/\asp$ on the range $\a \in (0, 2)$.  The informative prior on $\tau$, the timescale of shortest memory, was chosen after examining a host of improper priors  which all led to improper posteriors.  Based on scientific considerations, $\tau$ was set to have 99\% probability \emph{a priori} of falling between \mbox{\SI{e-6}{\second} and \SI{1}{\second}}.


The priors in~\eqref{eq:specprior} are members of the conjugate family
\begin{equation}\label{eq:conjprior}
\begin{split}
\ap & \sim \pi(\ap), \\
\SSigma \| \ap & \sim \tx{Inv-Wishart}(\bm{\Psi}_{\ap}, \nu_{\ap}) \\
\mmu \| \SSigma, \ap & \sim \N_2(\bm{\lambda}_{\ap}, \SSigma/\kappa_{\ap}),
\end{split}
\end{equation}
where $\bm{\Psi}_{\ap}$, $\nu_{\ap}$, $\bm{\l}_{\ap}$, and $\kappa_{\ap}$ can each depend on the value of $\ap$.  For this family of priors, the posterior distribution is
\begin{equation}\label{eq:conjpost}
\begin{split}
\ap \| \x & \sim \pi(\ap) \times \frac{\Lambda(\bm{\Psi}_{\ap}, \nu_{\ap})}{\Lambda(\bm{\hat \Psi}_{\ap}, \hat \nu_{\ap})} \times \frac{\kappa_{\ap}}{\hat \kappa_{\ap}\abs{V_{\ap}}} \\
\SSigma \| \ap, \x & \sim \textrm{Inv-Wishart}(\bm{\hat \Psi}_{\ap}, \hat \nu_{\ap}) \\
\mmu \| \SSigma, \ap, \x & \sim \N_2(\bm{\hat \l}_{\ap}, \SSigma/\hat \kappa_{\ap}),
\end{split}
\end{equation}
where expressions for $\Lambda(\cdot, \cdot)$, $\bm{\hat \Psi}_{\ap}$, $\hat \nu_{\ap}$, $\bm{\hat \l}_{\ap}$, and $\hat \kappa_{\ap}$ are provided in~\ref{sec:marginf}.  Thus we were able to compute our models' subdiffusion parameter posteriors $p(\ap \| \x)$ using one- and two-dimensional grids.

\begin{figure}[htb!]
	\centering
	\begin{subfigure}{\textwidth}
		\includegraphics[width=1.00\textwidth]{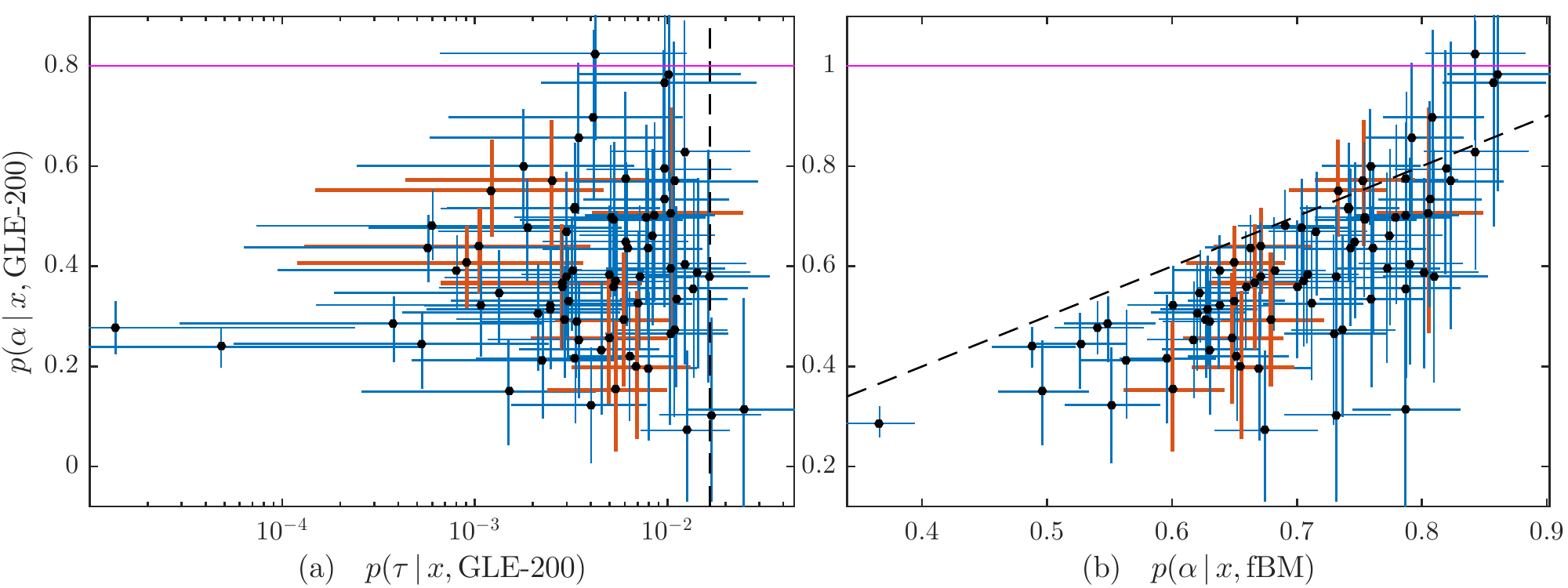}\phantomsubcaption\label{fig:alphaposta}\phantomsubcaption\label{fig:alphapostb}
	\end{subfigure}
	\caption{(a) Posterior mean and 95\% credible intervals for $\a$ and $\tau$ of the GLE-200 model. Red lines correspond to the 10 datasets in Figure~\ref{fig:data}.  The pink line indicates the subdiffusion threshold $\a < 1$. (b) Posterior mean and 95\% credible intervals for the subdiffusion coefficient $\alpha$ of the fBM and GLE-200 models.  The dashed \SI{45}{\degree} line indicates identical posterior means.}
	\label{fig:alphapost}
\end{figure}
Figure~\ref{fig:alphaposta} displays posterior means and 95\% credible intervals for the subdiffusion parameter $\a = 1/\asp$ and shortest memory parameter $\tau$ for the GLE model with $K = 200$ modes (hereafter GLE-200).  This choice of $K$ was made after very similar results were obtained with $K = 100$.  Figure~\ref{fig:alphaposta} reveals that $\tau$ is well below $\d t = 1/60$s, and thus difficult to measure precisely.  Higher frequency trajectories are therefore required to identify the timescale of shortest memory.


Figure~\ref{fig:alphapostb} displays posterior means and 95\% credible intervals for the subdiffusion parameters of the fBM and GLE-200 models: $\a_\tx{fBM} = 2H$ and $\a_\tx{GLE} = 1/\asp$. 
Both fBM and GLE-200 provide strong evidence against ordinary diffusion, with almost no credible intervals containing $\a = 1$.  However, GLE-200 provides consistently lower estimates of $\a$, except in the vicinity of $\a = 1$.  

\begin{figure}[htb!]
	\centering
		\includegraphics[width=1.00\textwidth]{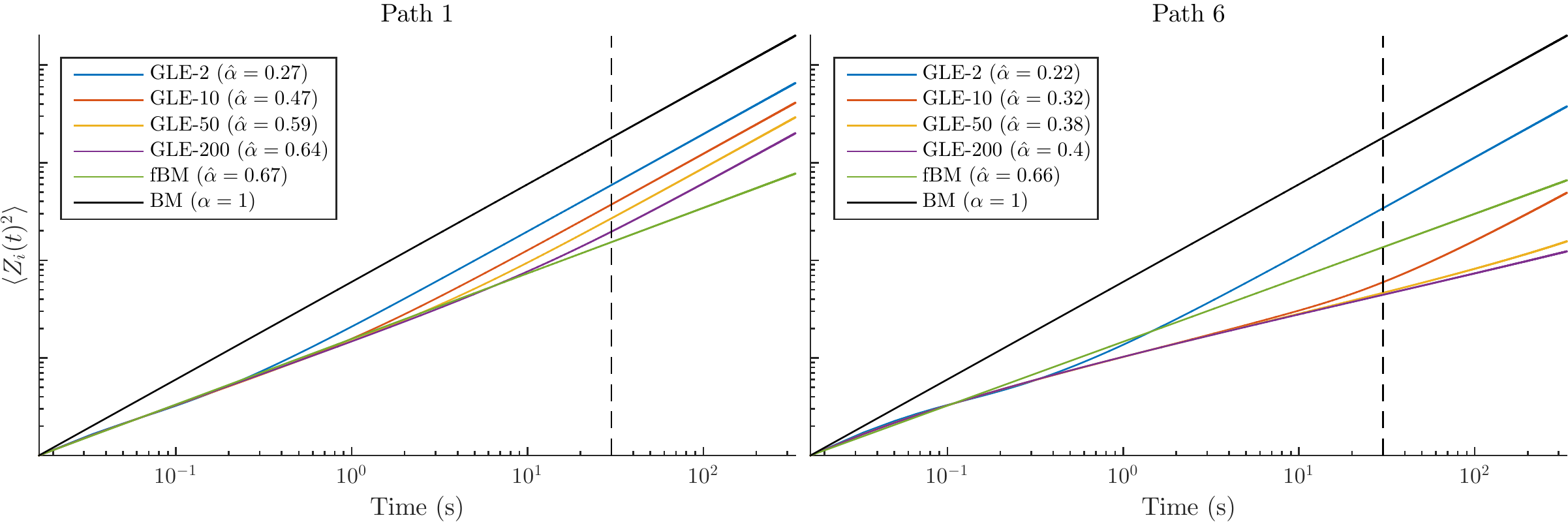}
	\caption{Theoretical MSDs for representative trajectories from Figure~\ref{fig:data}.  The dashed line demarquates the observation timeframe (0-30 seconds).}  
	\label{fig:msd}
\end{figure}

To investigate this pattern, Figure~\ref{fig:msd} displays theoretical MSDs for the subdiffusive process $\inner{Z^2_i(t)}$.  These MSDs are evaluated at the posterior parameter means of two representative particle trajectories, for fBM and GLE models with $K = 2, 10, 50,\tx{and}\ 200$ modes (recall that GLE-2 is the sum of Brownian motion and a single independent OU process).
All subdiffusive models report near-identical MSD slopes at short timescales.  As fBM has uniform subdiffusion, it extends this slope to all timescales (straight line on the log-log scale).  The GLEs on the left agree with fBM at intermediate timescales as well, effectively fitting $t_0 \approx 0$ for the onset of the subdiffusive range in~\eqref{eq:trans-subdiff}.  However, the GLEs on the right exhibit noticeably lower subdiffusion after an initial period, hence the lower estimate of $\a$ for $t \in (t_0,t_1)$.  We note that, aside from GLE-2, the transition to ordinary diffusion ($\a = 1$) occurs near or beyond the 30 second observation period.  We hypothesize that this is likely not a scientifically meaningful finding, but rather an artifact of short timescales driving the model fit.  Evidence to support this claim is offered in Section~\ref{sec:gof}.


\section{Bayesian Model Comparisons}\label{sec:modelcomp}

In the Bayesian framework we have adopted, a natural quantitative approach to model selection is through the use of Bayes factors~\citep{jeffreys61, kass-raftery95, green95}.  Bayes factors have been widely used to compare complex and non-nested models of biological systems \citep{lartillot-philippe06, vyshemirsky-girolami08, toni-et-al09, li-drummond12}, and recently, to analyze MSD curves of particle motion in live cells~\citep{monnier-et-al12}.  However, the strategy adopted by these last authors consists of least-squares fitting to empirical MSD curves.  As these are subject to considerable sampling variation, a great deal of selection power is lost by renouncing a fully parametric likelihood-based approach. Furthermore, the approach relies solely on summary statistics (the pathwise MSD and its variants), in which case a great deal of care must be employed to insure that Bayesian model selection leads to statistically valid results~\citep{robert-et-al11}.

Suppose that we wish to compare two models $M_1$ and $M_2$ which assign distributions $p(\x \| \t_i, M_i)$ to the observed data $\x$, where $\t_i$ denotes the parameter vector of model $M_i$.  Then for proper priors $\pi(\t_i \| M_i)$, the posterior probability of model $M_i$ is
\begin{equation}\label{eq:modpost}
p(M_i \| \x) = \frac{q_i f_i(\x)}{q_1 f_1(\x) + q_2 f_2(\x)},
\end{equation}
where $q_i$ is the prior probability for $M_i$ (such that $q_1 + q_2 = 1$), and
\[
f_i(\x) = p(\x \| M_i) = \int p(\x \| \t_i, M_i) \pi(\t_i \| M_i) \ud \t_i
\]
is the marginal likelihood under $M_i$.  The Bayes factor (BF) is defined as a ratio of posterior to prior model odds,
\[
\textrm{BF} = \frac{p(M_1 \| \x)/p(M_2 \| \x)}{q_1/q_2} = \frac{f_1(\x)}{f_2(\x)},
\]
which can be used to assess the relative plausibility of $M_1$ to $M_2$.  An attractive feature of this measure is that it does not depend on $q_i$.   Equivalently, the Bayes factor is simply the posterior odds $p(M_1 \| \x)/p(M_2 \| \x)$ when a lack of preference between either model is expressed as equal prior probabilities $q_1 = q_2 = \tfrac 1 2$.  In this case we have the monotone transformation $p(M_1 \| \x) = \tx{BF}/(1+\tx{BF})$, which we have used here to interpret the Bayes factor calibration on a probability scale.

A major criticism of the Bayesian model selection approach is that the choice of prior  distribution $\pi(\t_i \| M_i)$ can have a considerable impact on the posterior probabilities in~\eqref{eq:modpost} (see for instance~\citealp{lindley57, kass-raftery95, berger-pericchi96} and many other references in~\citealp{vanpaemel10}). For our particular application, two ``default'' priors are evaluated in~\ref{sec:badprior}: (i) a noninformative but proper prior, and (ii) the informative prior of~\cite{aitkin91} which uses the data twice.  Both are found to emphatically select the GLE-200 model over fBM, regardless of which model is used to simulate the data.

Fortunately, another default prior construction is available for the multiply replicated experiment at hand.  That is, priors for the 10 ``testing'' datasets retained for model comparison in Figure~\ref{fig:data} can be obtained by pooling the remaining 66 ``training'' datasets through the use of a hierarchical model.  Specifically, for each model $M_i$, the hierarchical model on all $J = 76$ datasets $\rv {\x} J$ is
\begin{equation}\label{eq:hm}
\x_j \| \t_j \ind f(\x_j \| \t_j), \qquad \t_j \iid g(\t \| \bm\eta), \qquad \bm\eta \sim \pi(\bm\eta),
\end{equation}
where we have dropped the dependence on $M_i$ to simplify notation.  Here, $f(\x \| \t)$ are the matrix-normal densities~\eqref{eq:lsincr}, $\t$ are the parameters of the fBM or GLE models, $\t_\tx{fBM} = (H, \mmu, \SSigma)$ and $\t_\tx{GLE} = (\asp, \tau, \mmu, \SSigma)$, and $\bm \eta$ are the hyperparameters of the hierarchical model.  Such a model naturally induces between-path particle heterogeneity and the ergodicity breaking phenomenon described in Section~\ref{sec:contrib}.

Fitting~\eqref{eq:hm} to the $T = 66$ training datasets $\x_{\tx{train}} = (\rv {\x} T)$ produces a posterior distribution
\begin{equation}\label{eq:trainpost}
p(\bm\eta \| \x_{\tx{train}}) 
\propto \pi(\bm\eta) \prod_{j=1}^T \int f(\x_j \| \t_j) g(\t_j \| \bm\eta) \ud \t_j,
\end{equation}
which leads to a proper parameter prior for the 10 testing datasets $\x_\tx{test} = (\rv [T+1] {\x} J)$ in Figure~\ref{fig:data}: 
\begin{equation}\label{pitest}
\pi_\tx{test}(\t) = p(\t \| \x_\tx{train}) = \int g(\t \| \bm\eta) p(\bm\eta \| \x_\tx{train}) \ud \bm\eta.
\end{equation}
This approach to prior specification for Bayesian model comparisons is a simplified instance of the \emph{intrinsic Bayes factor} of~\cite{berger-pericchi96}.  A full application of this procedure would average over all ``minimal'' training samples capable of identifying the model parameters.  In our context however the resulting computations become prohibitively expensive, and we proceed with the 66 training sets and 10 test sets from Figure~\ref{fig:data}.


\subsection{Approximate Fitting of the Hierarchical Model}\label{sec:hmapprox}

While the prior $\pi_\tx{test}(\t)$ in~\eqref{pitest} is conceptually appealing, its estimation typically requires computationally intensive Markov chain Monte Carlo (MCMC) techniques.  As an alternative, we outline here an approximation which allows the bulk of the calculations to be run in parallel.
\begin{enumerate}
\item Suppose that $\t$ is a $d$-dimensional parameter vector.  We begin by obtaining posterior samples from
\begin{equation}\label{eq:indinf}
p_0(\t_j \| \x_j) = f(\x_j \| \t_j)g_0(\t_j)/C_j, \qquad C_j = \int f(\x_j \| \t_j)g_0(\t_j) \ud \t_j,
\end{equation}
for each training set $\rv {\x} T$.  Any prior $g_0(\t)$ can be used, although the approximation works best when $g_0(\t)$ is uninformative.  
In our case, $g_0(\t)$ is the improper prior~\eqref{eq:specprior} defined in Section~\ref{sec:fit}.  This stage can be implemented in parallel as the inference for each dataset is independent from any other.
\item\label{it:normapprox} 
Suppose that each posterior $p_0(\t_j \| \x_j)$ is approximately normal:
\begin{equation}\label{eq:normapprox}
p_0(\t_j \| \x_j)
\approx \varphi(\t_j \| \l_j, \OO_j),
\end{equation}
where $\varphi(\cdot \| \l, \OO)$ is the PDF of a Gaussian distribution with mean $\l$ and variance $\OO$.  Then if \emph{a priori} we have
\[
\t_j \iid \N(\l_0, \OO_0),
\]
and $\pi(\bm \eta)$ is a prior on $\bm \eta = (\l_0, \OO_0)$, the posterior distribution on $\bm\Theta = (\rv \t T)$ and $\bm \eta$ is approximately
\begin{equation}\label{hmapprox}
\begin{split}
p(\bm\Theta, \bm \eta \| \x_\tx{train}) \propto\ & \pi(\bm \eta) \prod_{j=1}^T \varphi(\t_j \| \bm \eta) \times \big[f(\x_j \| \t_j)/C_j\big] \\
\approx\ & \pi(\bm \eta) \prod_{j=1}^T \varphi(\t_j \| \bm \eta) \times \big[\varphi(\t_j \| \l_j, \OO_j)/ g_0(\t_j)\big].
\end{split}
\end{equation}
\item\label{it:gibbs} For the Lebesgue prior $g_0(\t) \propto 1$, the approximate posterior~\eqref{hmapprox} has precisely the form of a multilevel normal model, for which Bayesian inference can easily be conducted with the help of a Gibbs sampler.  That is, for the scale-invariant hyperparameter prior
\[
\pi(\l_0, \OO_0) \propto \abs {\OO_0}^{-(\omega+d+1)/2},
\]
the Gibbs sampler updates its various components using the analytical distributions
\begin{equation}\label{eq:gibbs}
\begin{split}
\t_j \| \l_0, \OO_0, \x_\tx{train} & \ind \N_d\left(\bm B_j \l_0 + (\bm I-\bm B_j) \l_j, (\bm I-\bm B_j) \OO_j \right) \\
\OO_0 \| \t, \x_\tx{train} & \sim \tx{Inv-Wishart}\left(\bm S, \omega\right) \\
\l_0 \| \t, \OO_0, \x_\tx{train} & \sim \N_d\left(\bar {\l}, \tfrac 1 t \OO_0 \right),
\end{split}
\end{equation}
where $\bm B_j = \OO_j (\OO_j + \OO_0)^{-1}$, $\bar {\l} = \frac 1 t \sum_{j=1}^t \l_j$, and $\bm S = \sum_{j=1}^t (\l_j - \bar {\l})(\l_j - \bar {\l})'$.  When $g_0(\t)$ is not Lebesgue, a Metropolis step is used to correct the conditional draws of $\t_j$ with acceptance rate $a = \min\big\{1, g_0(\t_j^{(\tx{old})})/g_0(\t_j^{(\tx{new})})\big\}$.
\item To obtain samples from $\pi_\tx{test}(\t)$ in~\eqref{pitest}, each MCMC draw from $p(\l_0, \OO_0 \| \x_\tx{train})$ in step~\ref{it:gibbs} is augmented with a draw from $\t \| \l_0, \OO_0 \sim \N\left(\l_0, \OO_0\right)$.
\item Finally, the MCMC samples $(\t^{(1)}, \ldots, \t^{(m)})$ from $\pi_\tx{test}(\t)$ are used to approximate $\pi_\tx{test}(\t)$ by a prior in the conjugate family~\eqref{eq:conjprior} (as described in~\ref{sec:marginf}).  This proved to be a minor adjustment in our case, but with considerable computational benefits.  This is because model comparisons rely on the marginal likelihood 
$p(\x \| M_i) = \int p(\x \| \t_i, M_i) \pi_\tx{test}(\t_i \| M_i) \ud \t_i$.  Under a conjugate prior, $p(\x \| M_i)$ is efficiently calculated for fBM and GLE models via one- and two-dimensional deterministic integrals (\ref{sec:marginf}).
\end{enumerate}

\begin{figure}[htb!]
	\centering
		\includegraphics[width=1.00\textwidth]{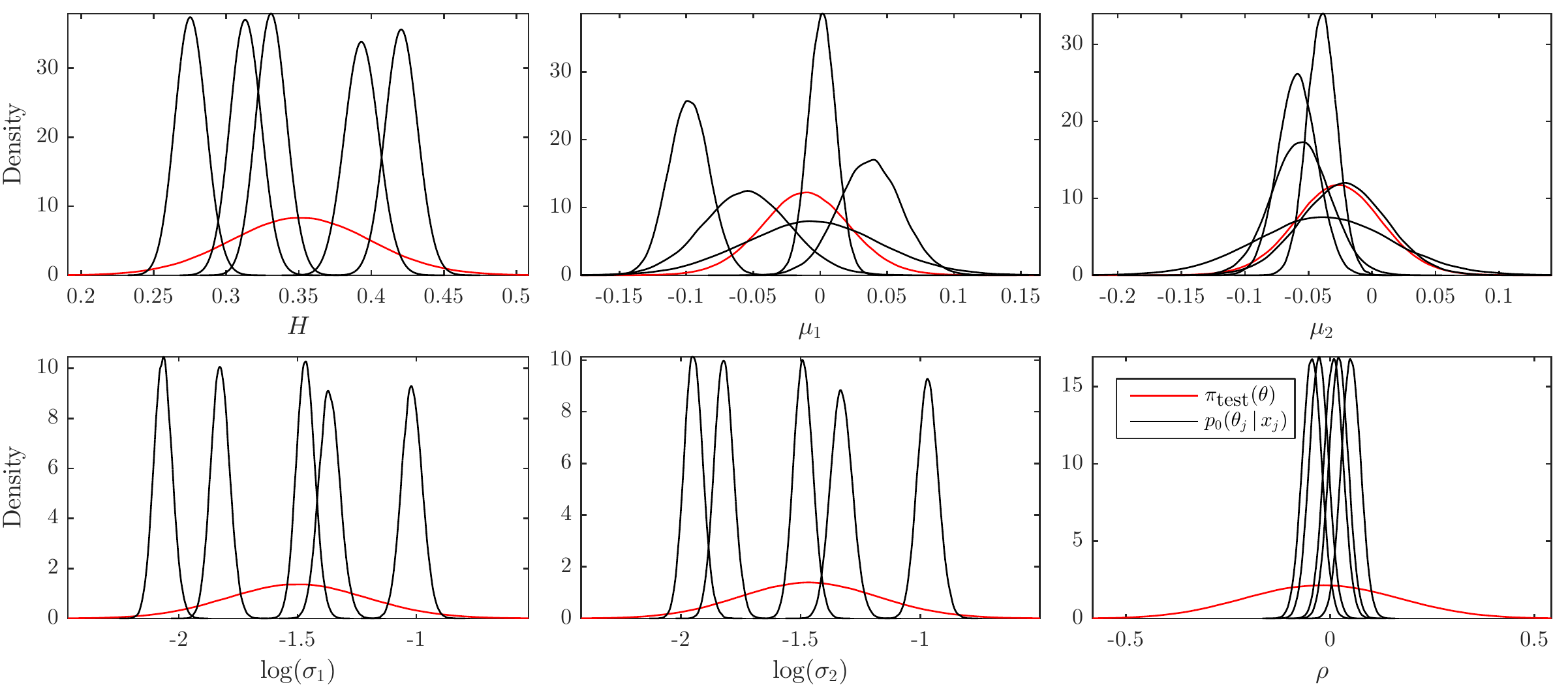}\\
		\includegraphics[width=1.00\textwidth]{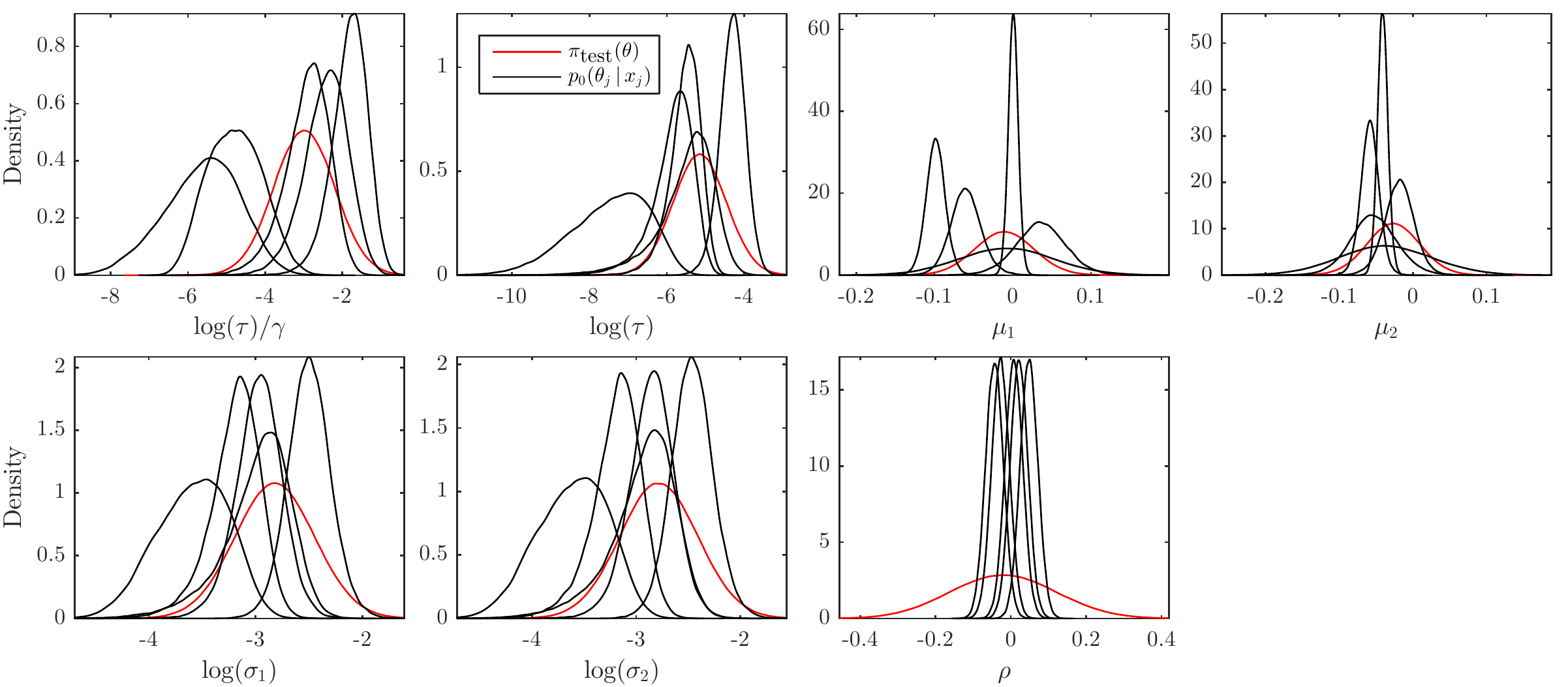}
	\caption{Testing priors (red) and independent posteriors (black) for the fBM and GLE-200 models.}
	\label{fig:train}
\end{figure}

An attractive feature of the approximate hierarchical model above is that the bulk of its calculations -- fitting each $p_0(\t_j \| \x_j)$ -- can be conducted in parallel.  A similar approach was adopted by~\cite{lunn-et-al13} to fit the hierarchical model exactly, employing a Metropolis-Hastings resampling step for the conditional draws from $p(\t_j \| \bm \eta, \x_\tx{train})$.  A potential drawback of this exact MCMC algorithm is that the Metropolis-Hastings acceptance rates are adversely affected when the independent posteriors are shrunk heavily towards the common mean (as for $\mu_1$ and $\mu_2$ in Figure~\ref{fig:train}).

To improve the accuracy of the normal approximation~\eqref{eq:normapprox}, we transformed the parameters to
\begin{align*}
\t_\tx{fBM} & = \Big(H, \mu_1, \mu_2, \log(\sigma_1), \log(\sigma_2), \rho\Big), \\
\t_\tx{GLE} & = \Big(\log(\tau)/\asp, \log(\tau), \mu_1, \mu_2, \log(\sigma_1), \log(\sigma_2), \rho\Big).
\end{align*}
Figure~\ref{fig:train} displays the densities of $\pi_\tx{test}(\t)$ for the fBM and GLE-200 models under the normalizing transformations, along with a sample of the independent posteriors $p_0(\t_j \| \x_j)$ which were used to compute them.

\subsection{Model Comparison Results}

To evaluate the adequacy for model selection of the data-driven testing prior~\eqref{pitest}, 500 datasets from fBM and GLEs with $K =2, 10, 50, 200$ were simulated with parameters drawn from $\pi_\tx{test}(\t_i \| M_i)$.  Posterior model probabilities~\eqref{eq:modpost} were calculated for each dataset and two-way model comparison.  Simulation results are summarized in Table~\ref{tab:probsel}.
\settowidth{\colw}{GLE-200}
\begin{table}[!htb]
	\centering
\begin{tabular}{crWWWWW}
\multicolumn{2}{c}{} & \multicolumn{5}{c}{Alternative Model} \\
\multicolumn{2}{c}{} & GLE-2 & GLE-10 & GLE-50 & GLE-200 & fBM \\
\cline{3-7}
\multirow{5}{*}{\rotatebox[origin=c]{90}{Correct Model}}
& GLE-2 & - & 88 (92) & 90 (93) & 91 (94) & 98 (100) \\
& GLE-10 & 87 (92) & - & 59 (64) & 62 (68) & 87 (90) \\
& GLE-50 & 90 (93) & 56 (69) & - & 53 (61) & 88 (92) \\
& GLE-200 & 93 (96) & 64 (78) & 51 (59) & - & 81 (84) \\
& fBM & 96 (97) & 85 (92) & 84 (90) & 83 (90) & - \\
\cline{3-7}
\end{tabular}
	\caption{Summary of model selection results with simulated data.  The first number corresponds to the average posterior probability $p(M_c \| \x) \times 100\%$ attributed to the correct model $M_C$ under equal prior odds.  The number in parentheses is the percentage of datasets for which the correct model had larger posterior odds.}
	\label{tab:probsel}
\end{table}
For each pair of models, the first number is the average posterior probability $p(M_C \| \x)$ given to the correct model $M_C$, taken over datasets generated from the correct model, $\x \sim p(\x \| M_C)$.  The number in parentheses is the probability of selecting the correct model $M_C$, based on the rule which chooses the model with the highest posterior probability.  Table~\ref{tab:probsel} indicates that fBM, GLE-2, and GLE-200 are highly distinguishable from each other using Bayes factors, with 80-95\% classification accuracy.  On the other hand, the intermediate GLE models with $K = 10,50,200$ are more difficult to tell apart (60-80\% accuracy).  This is consistent with the relatively similar MSD estimates reported by these models over the experimental timescale in Figure~\ref{fig:msd}.

\begin{figure}[htb!]
	\centering
		\includegraphics[width=1.00\textwidth]{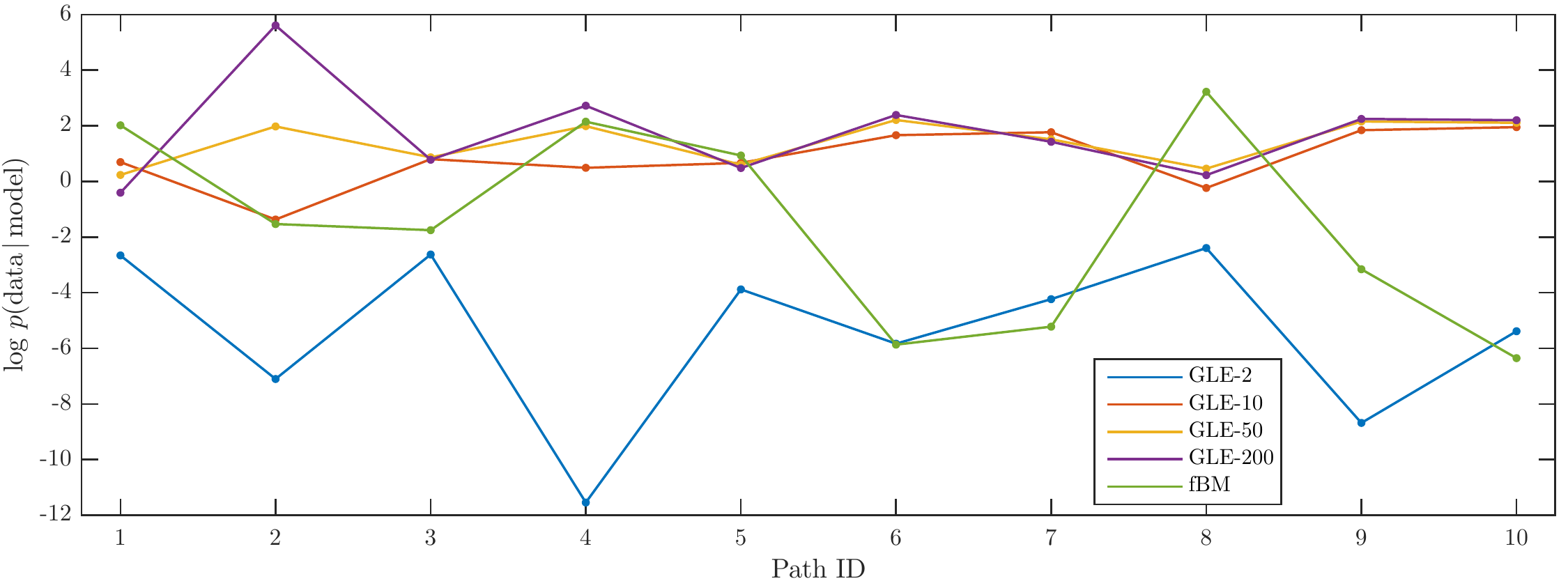}
	\caption{Marginal likelihoods for fBM and GLE models computed for the 10 particle trajectories in Figure~\ref{fig:data}.}
	\label{fig:bf}
\end{figure}

We now turn to model selection for the ten real datasets of Figure~\ref{fig:data}.  The marginal likelihoods $p(\x_j \| M_i)$ for each dataset and model are displayed in Figure~\ref{fig:bf}.
Within each dataset, the marginal likelihoods for all models were multiplied by a constant to facilitate visualization, such that comparisons across datasets cannot be made.  However, model comparisons within each dataset suggest that the GLEs have marginal likelihood increasing with $K$.

Table~\ref{tab:probdata} displays the posterior probability of the best-fitting GLE model against the fBM alternative.
\begin{table}[htb!]
	\centering
\begin{tabular}{r|cccccccccc}
\hline Dataset &  1  &  2  &  3  &  4  &  5  &  6  &  7  &  8  &  9  &  10  \\ \hline \hline
$\tx{Pr}(\tx{GLE-}K \| \x)$ (\%) &  21.0  &  99.9  &  93.2  &  63.9  &  43.3  &  99.9  &  99.9  &  5.9  &  99.5  &  99.9  \\
$K$ &  10  &  200  &  50  &  200  &  10  &  200  &  10  &  50  &  200  &  200  \\ \hline
\end{tabular}
\caption{Posterior probability ($\times 100\%$) of GLE model vs.\ fBM for each of the ten test sets from Figure~\ref{fig:data} (top row).  For each dataset the GLE with the most favorable number of modes was used (bottom row).}\label{tab:probdata}
\end{table}
In six of the ten test sets, the best-fitting GLE emphatically dominates fBM (over 90\% posterior probability), with the reverse pattern only exhibited in dataset DS-8.  In half the datasets, the best-fitting GLE is GLE-200, and in most cases, Figure~\ref{fig:bf} shows that the worst-fitting model is GLE-2. 
These findings suggest that the GLE with dense relaxation spectrum (moderate to large $K$) most adequately describes the non-uniform subdiffusion exhibited by our data.


\section{Predictive Model Assessment of Goodness-of-Fit}\label{sec:gof}

In this section we wish to assess whether the observed particle trajectories are consistent with our candidate subdiffusive models.  To do this, we begin with the null hypothesis that the data come from a hierarchical model:
\begin{equation}\label{eq:h0}
H_0: \x \| \t \sim f(\x \| \t), \quad \t \sim \pi_\tx{test}(\t).
\end{equation}
Here, $\x = \x_\tx{test}$ is one of the ten datasets of Figure~\ref{fig:data}, $f(\x \| \t)$ is the matrix-normal density~\eqref{eq:lsincr} of the fBM or GLE increments, and $\pi_\tx{test}(\t) = p(\t \| \x_\tx{train})$ is obtained from the 66 training datasets in Section~\ref{sec:modelcomp}.  While this prior is data-driven, it does not depend on $\x_\tx{test}$.  Therefore, $\x_\tx{test} \sim p(\x \| H_0)$ has a sampling distribution which can be interpreted in the Frequentist sense, if one accepts both levels of the hierarchical model.

To evaluate various aspects of the model under $H_0$, we consider a test statistic $T = T(\x)$, and compare its observed value $T_\obs = T(\x_\obs)$ to either a \emph{prior predictive distribution}~\citep{box80}
\begin{equation}\label{eq:pprior}
p_\tx{prior}(T) = \int p(T \| \t, H_0) \pi_\tx{test}(\t) \ud \t,
\end{equation}
or a \emph{posterior predictive distribution}~\citep{rubin84}
\begin{equation}\label{eq:ppost}
p_\tx{post}(T) = \int p(T \| \t, H_0) p(\t \| \x_\obs, H_0) \ud \t.
\end{equation}
Lack of concordance between $\x_\obs$ and the model are indicated by $T_\obs$ in the tails of~\eqref{eq:pprior} or~\eqref{eq:ppost}.  The choice of which predictive distribution to use depends on the model feature under examination, and shall be made explicit in the following subsections.

We employ two sets of test statistics to evaluate our subdiffusive models.  The first pertains to the MSD -- the model feature of primary interest (Subsection~\ref{sec:msdtest}).  None of our MSD tests exhibit evidence of model inadequacy, though we note that they have limited power to detect anomalies at longer timescales.  The second set of tests is based on conditionally Gaussian residuals, which are both easy to calculate and extremely sensitive to model miss-specifications (Subsection~\ref{sec:residtest}).  In the worst-fitted datasets, these residuals reveal a pattern of within-path heterogeneity which our models do not capture.


\subsection{MSD-Based Tests}\label{sec:msdtest}

\begin{figure}[htb!]
	\centering
		\includegraphics[width=1.00\textwidth]{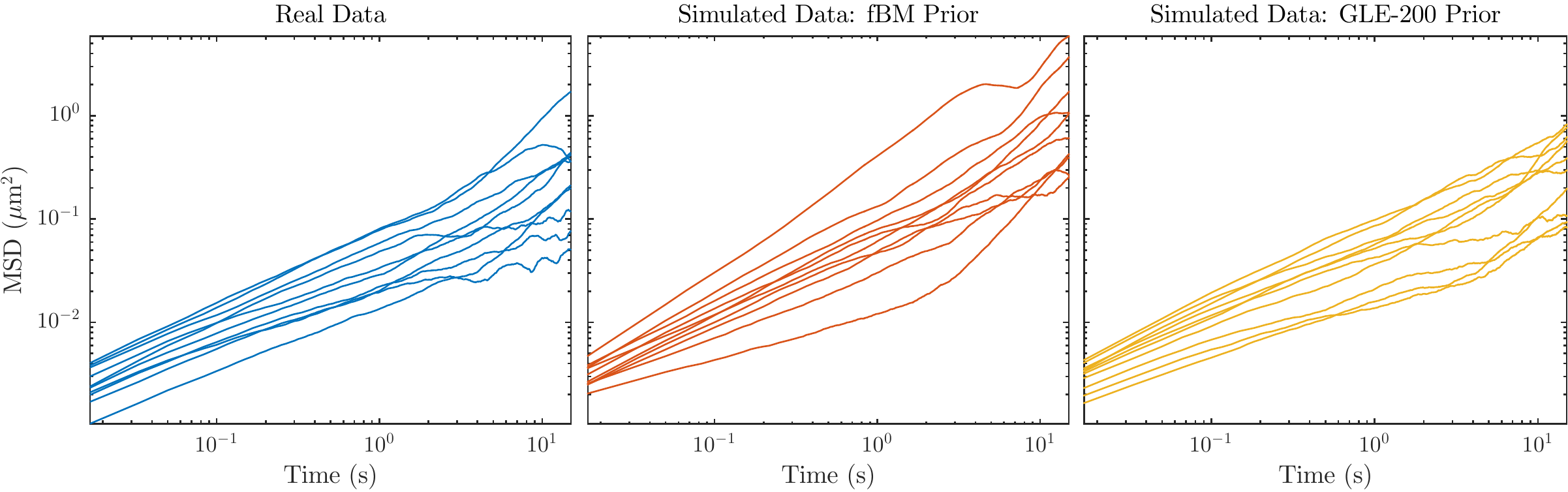}
	\caption{Sample MSDs for empirical data and simulated data from the prior predictive distribution.}\label{fig:ergbreak}
\end{figure}

We have seen in Figure~\ref{fig:heterdata} that the 76 experimental particle trajectories exhibit considerably more between-path heterogeneity than a single-parameter fBM or GLE model would predict.  Figure~\ref{fig:ergbreak} compares the pathwise MSD statistics for the ten real datasets of Figure~\ref{fig:data}, $\x_\obs^{(i)}$, $i = 1,\ldots,10$, to those of ten simulated datasets from the prior predictive distribution~\eqref{eq:pprior} under fBM and GLE-200 hierarchical models. For two-dimensional paths, the sample MSD was calculated as the average MSD statistic~\eqref{eq:defn-pathwise-msd} for each one-dimensional trajectory, after ``detrending'' the sample mean of the increments to zero. 
The MSDs in Figure~\ref{fig:ergbreak} are similar to those of the empirical data.  This demonstrates that a hierarchical fBM or GLE model can induce the observed amount of ergodicity breaking -- a central feature of the CTRW model discussed in Section~\ref{sec:intro}.

\begin{figure}[htb!]
	\centering
		\includegraphics[width=1.00\textwidth]{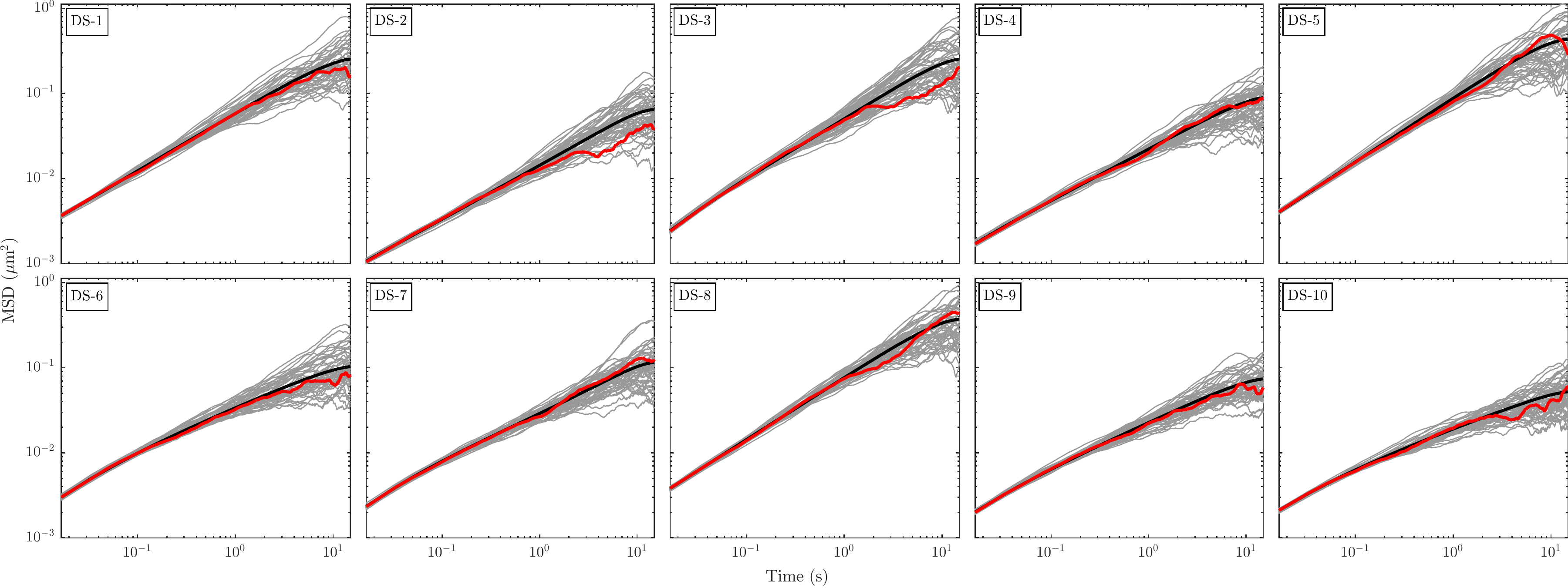}
	\caption{Simulated MSD statistics from their posterior predictive distributions.  Observed MSD (red) and average posterior MSD (black) and are superimposed.}\label{fig:msdpred}
\end{figure}

Figure~\ref{fig:msdpred} compares the MSD statistic of each $\x_\obs^{(i)}$ to 100 simulated trajectories from its posterior predictive distribution, under the model selected by the Bayes factor in Section~\ref{sec:modelcomp} (fBM: 1,5,8; GLE-200: 2,4,6,9,10; GLE-50: 3; GLE-10: 7).
Also provided in Table~\ref{tab:msdpval} are posterior predictive $p$-values,
\[
\pvpost = \tx{Pr}_\tx{post}(T > T_\obs),
\]
a quantitative measure of the lack of concordance between simulated and observed MSD statistics.
\settowidth{\colw}{100}
\begin{table}[!htb]
	\centering
\centerline{
\begin{tabular}{cr|WWWWWWWWWW}
\multicolumn{2}{c}{} & \multicolumn{10}{c}{Dataset} \\
\multicolumn{2}{c|}{Statistic} & 1 & 2 & 3 & 4 & 5 & 6 & 7 & 8 & 9 & 10 \\
\hline
\multirow{4}{*}{\rotatebox[origin=c]{90}{$T = \widehat{\tx{MSD}}(t)$}}
& $t=\sfrac{1}{60}$ s & 57 & 32 & 38 & 43 & 55 & 56 & 46 & 56 & 41 & 48 \\
& $t=\sfrac{1}{10}$ s & 22 & 59 & 45 & 59 & 47 & 51 & 41 & 33 & 58 & 28 \\
& $t=1.0$ s & 52 & 26 & 46 & 28 & 33 & 41 & 29 & 44 & 46 & 58 \\
& $t=10$ s & 43 & 25 & 18 & 54 & 76 & 24 & 76 & 68 & 45 & 44 \\
\hline
\end{tabular}
}
	\caption{Posterior predictive $p$-values ($\times 100\%$) for the test statistic $T = \widehat{\tx{MSD}}(t)$ at different values of $t$.}
	\label{tab:msdpval}
\end{table}

Under $H_0$, the posterior predictive distribution draws a new dataset $\x$ from the \emph{same} unknown value of $\t$ which generated $\x_\obs$, and thus is well-suited for assessing within-path particle dynamics.  Unlike classical $p$-values, the distribution of $\pvpost = \pvpost(\x_\obs)$ under $\x_\obs \sim p(\x \| H_0)$ is generally non-uniform, tending to be conservative of $H_0$~\citep{meng94}.  Nonetheless, $\pvpost$ remains a valid probability under $H_0$, with the usual $p$-value interpretation: that of $T = T(\x)$ for the new dataset providing more evidence against $H_0$ than $T_\obs$~\citep{gelman13}. 

At short timescales (\numrange[range-phrase=--]{2}{5} s), the simulated MSDs in Figure~\ref{fig:msdpred} and Table~\ref{tab:msdpval} are in excellent agreement with those of the empirical data. 
At longer timescales, the MSD statistic becomes highly variable, having little power to detect model departures.  We take these findings as indicative of the timescale over which the data provide reliable information.  In particular, we have little evidence to assess the GLE's transition to ordinary diffusion in Figure~\ref{fig:msd}.

\subsection{Analysis of Residuals}\label{sec:residtest}

If $\x_{N\times 2}$ has the matrix-normal distribution $f(\x \| \t)$ in~\eqref{eq:lsincr}, then
\begin{equation}\label{eq:zresid}
\Z_{N\times 2} = \Z(\x, \t) = \bm V_{\ap}^{-1/2} (\x - \bm{\Delta t}\mmu) \SSigma^{-1/2}
\end{equation}
is an $N\times 2$ matrix of iid $\N(0,1)$ residuals.  For $N = 1800$ observations, these residuals could be used to construct highly sensitive tests against the conditional model $f(\x \| \t)$.  As $\t$ is unknown, $\Z$ cannot be observed directly.  However, we may easily obtain draws from the conditional distribution
\begin{equation}\label{eq:ztest}
p(\Z \| \x_\obs) = \int p\big(\Z(\x_\obs, \t) \| \t\big) p(\t \| \x_\obs) \ud \t,
\end{equation}
by calculating $\Z(\x_\obs, \t)$ from a posterior draw of $\t \sim p(\t \| \x_\obs)$.  

Model evaluations with parameter-dependent test statistics such as~\eqref{eq:zresid} fall under the framework of ``realized discrepancy assessments'' proposed by~\cite{gelman-et-al96}, and similar residuals have been proposed by~\cite{albert-chib95} for outlier detection in regressions with binary outcomes.  Note that a draw from $\x_\obs \sim p(\x \| H_0)$ followed by a draw from $\Z \sim p(\Z \| \x_\obs)$ results in $\Z$ being a matrix of iid standard normals.  Moreover, a test statistic $T(\Z) = T(\x,\t)$ is amenable to posterior predictive $p$-value calculations of the form
\begin{align*}
\pvpost & = \tx{Pr}_\tx{post}\big(T(\x, \t) > T(\x_\obs, \t)\big) \\
& = \int \tx{Pr}\big(T(\x, \t) > T(\x_\obs, \t) \| \x_\obs, \t, H_0\big) p(\t \| \x_\obs, H_0) \ud \t.
\end{align*}

\begin{figure}[htb!]
	\centering
		\includegraphics[width=1.00\textwidth]{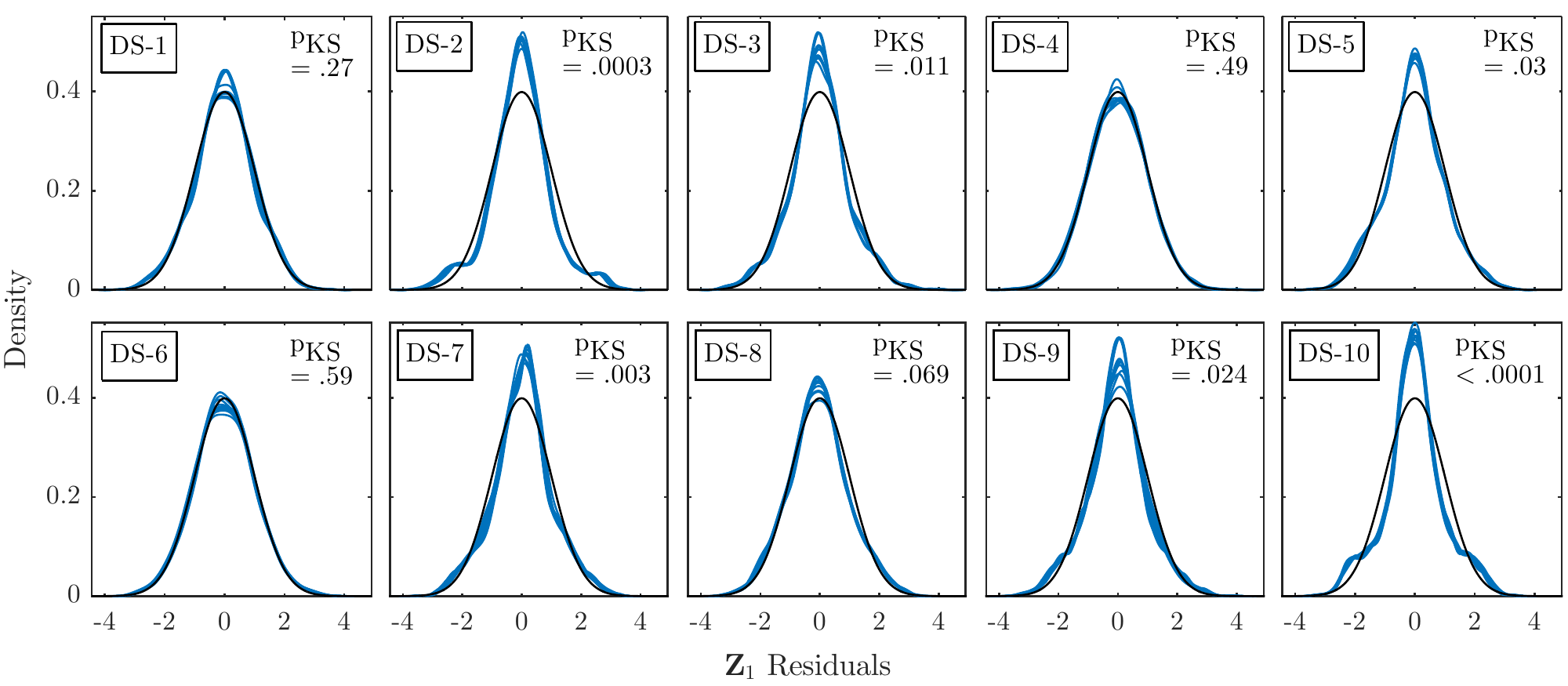}
		\includegraphics[width=1.00\textwidth]{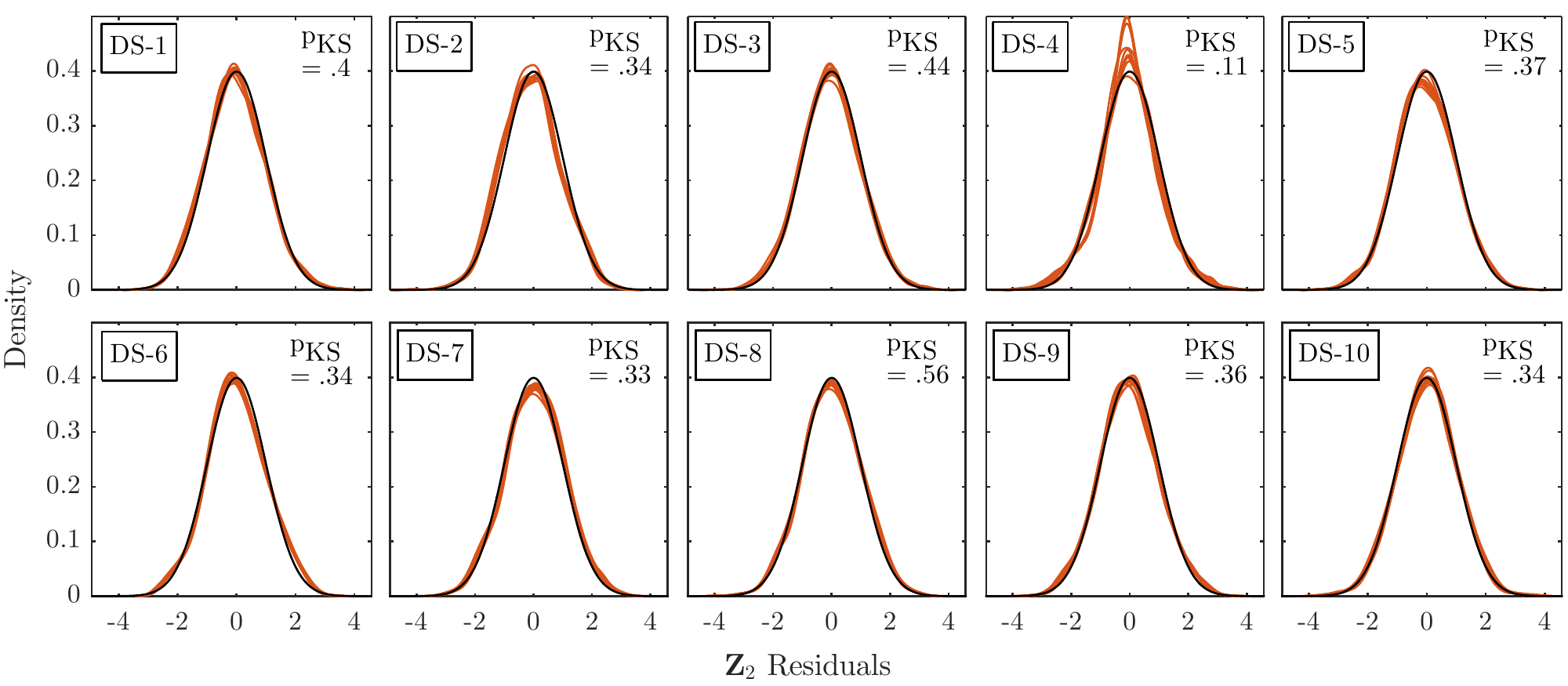}
	\caption{Bayesian model residuals $\Z = \Z(\x_\obs, \t)$ drawn from $p(\Z \| \x_\obs)$.  The columns of $\Z = (\Z_1, \Z_2)$ correspond to the major and minor eigenvectors of $\SSigma$.  Posterior predictive $p$-values for the Kolmogorov-Smirnov test are labeled $\tx{p}_\tx{KS}$.}\label{fig:pcresid}
\end{figure}

Figure~\ref{fig:pcresid} displays density estimates for ten draws from $p(\Z \| \x_\obs)$ for each dataset $\x_\obs^{(i)}$.  The square-roots of $\bm V_{\ap}$ and $\SSigma$ we employed were the Cholesky decomposition of the former, and the eigendecomposition of the latter.  Thus, for each $\t \sim p(\t \| \x_\obs)$, the elements of $\Z = (\Z_1, \Z_2)$ are of the form
\[
z_{nk} = \frac{\tilde x_{nk} - \E{\tilde x_{nk} \| \rv [1k] {\tilde x} {n-1,k}, \t}}{\tx{sd}(\tilde x_{nk} \| \rv [1k] {\tilde x} {n-1,k}, \t)},
\]
where $\Z_k = (\rv [1k] z {Nk})$ and $\tilde \x_n = (\tilde x_{n1}, \tilde x_{n2})$ are the projections of $\x_n$ onto the eigenvectors of $\SSigma$.  Thus, $\Z_1$ is along the direction of greatest particle movement, after removing the drift induced by the slowly-moving mucing environment.  Also included in Figure~\ref{fig:pcresid} are posterior predictive $p$-values for the Kolmogorov-Smirnov test statistic.

\begin{figure}[htb!]
	\centering
		\includegraphics[width=1.00\textwidth]{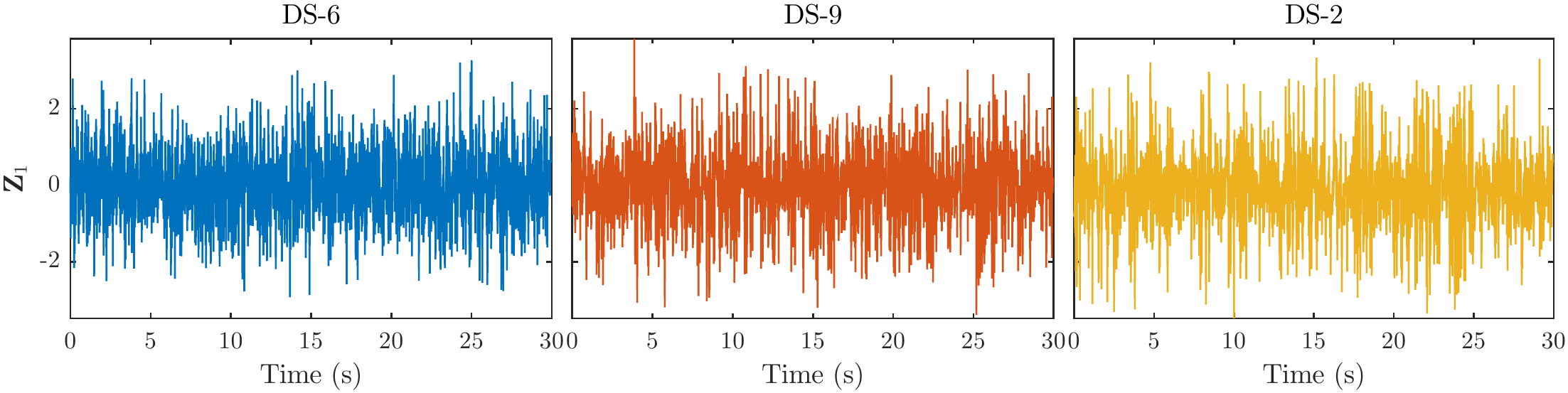}
	\caption{One posterior draw from $p(\Z_1 \| \x_\obs)$ for three representative datasets.}\label{fig:pc1ts}
\end{figure}

Figure~\ref{fig:pcresid} uncovers some lack-of-fit in the $\Z_1$ residuals.  
These appear to have a mixture distribution, most clearly visible in datasets DS-2 and DS-10.  To investigate serial dependence, Figure~\ref{fig:pc1ts} displays a single draw from $p(\Z_1 \| \x_\obs)$ for three datasets, each with increasingly worse model fit.
While the residuals for \mbox{DS-6} and DS-9 have no discernable temporal pattern, those of the severely misfit DS-2 have visible within-path heterogeneity, exhibited by alternating periods of large and small particle movement.  Interestingly, there is virtually no lack-of-fit in the $\Z_2$ residuals, despite the difference in variance along each eigenvector being only 25-50\%.


\section{Discussion}\label{sec:disc}


In this article we have detailed a rigorous Bayesian analysis of two models for subdiffusive two-dimensional particle trajectories in biological fluids.  The models are fBM and a GLE with generalized Rouse relaxation spectrum, respectively characterizing uniform and transient subdiffusive behaviors.  Our analyses leverage all information contained in the likelihood function, obviating the need for sole reliance on summary statistics such as the MSD.  Methodological contributions include a highly parallelizable approximation to the parameter posteriors of hierarchical models, and the development of versatile and easily computable residuals for conditionally Gaussian models.  Both of these methodologies are widely applicable outside the present setting.


Our analyses soundly suggest that the GLE with tunable power-law relaxation spectrum best describes the non-uniform subdiffusion exhibited by our data, on the timescale for which they provide reliable information.  The divergence between observed and expected MSD in Figure~\ref{fig:msdpred} suggests that this is about \numrange[range-phrase=--]{2}{5} s.  We have also found that a hierarchical fBM or GLE model naturally and adequately captures between-path heterogeneity of particle movement.
However, our highly sensitive residual analysis reveals some degree of model lack-of-fit.  In the worst-fit datasets, this was due to within-path particle fluctuations which our models do not resolve.  This hints at the presence of caging events, whereby particles move in and out of low-mobility cages formed by the complex molecular microstructure.

Based on these findings, we outline two important directions for further research.  The first pertains to modeling within-path heterogeneity.  One approach is to account for alternating periods of normal and confined particle movement via regime-switching models~\citep[e.g.][]{kim-nelson99}, the approach taken by CTRW.  
Along similar lines, one might allow the parameters of a ``homogenenous'' model such as fBM or GLE to depend on particle location.  By simultaneously examining multiple tracer particles in neighboring environments, such models could provide valuable information about the heterogeneity of the fluid sample.





A second direction for future work relates to the stated scientific objective of first-passage time prediction.  Experiments suggest that the passage time of several important pathogens through protective mucosal layers is on the order of dozens of minutes to hours~\citep{lai-et-al09c,suk-et-al11}.  This is far beyond the timescale over which we have reliable information on particle dynamics.  
Thus, to empirically assess the validity of first-passage time predictions, it will be necessary to increase the observation period by at least an order of magnitude.  However, this is difficult to achieve with the current experimental setup, as the tracer particles move out of the microscope focal plane given a sufficient amount of time.  It is possible to restrict attention to only the particles which remain in range for the duration of the experiment, but then inference must be conducted by conditioning on this fact.  We are actively looking into this approach, as well as different experimental particle-tracking strategies.


\newpage
\renewcommand{\appendixpagename}{\center \LARGE Supplementary Material\par\vskip 1em}

\begin{appendices}
\renewcommand{\thesection}{S\arabic{section}}
\setcounter{equation}{0}
\renewcommand{\theequation}{S\arabic{equation}}
\setcounter{figure}{0}
\renewcommand{\thefigure}{S\arabic{figure}}
\setcounter{table}{0}
\renewcommand{\thetable}{S\arabic{table}}

\section{Explicit Representation of the GLE}\label{sec:glecov}

Let $X(t)$ be the trajectory of the particle at time $t$.  A Generalized Langeving Equation (GLE) for $X(t)$ takes the form
\begin{equation}\label{eq:gle-const}
m \ddot X(t) = -\int_{-\infty}^t \phi(t-s) \dot X(s) \ud s + \sqrt{\kbt} F_t,
\end{equation}
where $m$ is the mass of the particle, $\dot X(t)$ and $\ddot X(t)$ are its velocity and acceleration, $\phi(t)$ is the memory kernel, $T$ is temperature, $k_B$ is Boltzmann's constant, and $F_t$ is a mean-zero stationary Gaussian process with autocorrelation
\[
\cov(F_t, F_s) = \phi(\abs{t-s}).
\]
We consider a sum-of-exponential memory kernel of the form
\[
\phi_K(t) = \eta \sum_{k=1}^K \exp(-\abs t \a_k).
\]
In the zero-mass limit $m\to 0$, applicable when ultra-high frequency particle dynamics have neglible impact, \cite{mckinley-et-al09} have shown that $X(t)$ with this particular kernel can be expressed as a linear combination of a Brownian motion $B_0(t)$ and $K-1$ Ornstein-Uhlenbeck processes
\[
\ud Y_{j}(t) = -r_j Y_{j}(t) \ud t + \ud B_{j}(t),
\]
all independent of each other.  That is, let $q(y) = \prod_{k=1}^K (y-\a_k)$, and $\rv r {K-1}$ be the roots of $q'(y)$.  Then
\begin{equation}\label{eqourep}
X(t) = \sqrt{\frac{2k_BT}{\eta}} \left(C_0 B_0(t) + \sum_{j = 1}^{K-1} C_j Y_{j}(t)\right),
\end{equation}
where 
\begin{align*}
C_0 & = \left(\sum_{k=1}^K \a_k^{-1}\right)^{-1/2}, & C_j & = \frac{1}{r_j} \times \frac{\sqrt{\sum_{k=1}^K \frac{\a_k}{(\a_k-r_j)^2}}}{\left(\sum_{k=1}^K \frac{1}{\a_k-r_j}\right)^2-\sum_{k=1}^K\frac{1}{(\a_k-r_j)^2}}.
\end{align*}

Using the representation~\eqref{eqourep}, the autocorrelation of the GLE increments $x_n = X(n \d t + \d t)-X(n \d t)$ is found to be
\begin{equation}\label{exacftrue}
\cov(x_i, x_j) = \frac{2k_BT}{\eta} \left(C_0^2 \d t \delta_k + \sum_{j=1}^{K-1} \frac{C_j^2}{2 r_j} \left(2e^{-r_j \abs k \d t} - e^{-r_j\abs{k-1}\d t} - e^{-r_j\abs{k+1}\d t}\right) \right),
\end{equation}
with $k = \abs{i-j}$ and $\delta_k$ the Dirac $\delta$-function.  Similarly, the MSD of the particle described by~\eqref{eq:gle-const} is
\begin{equation}\label{eq:exactmsd}
\inner{X(t)^2} = \frac{2k_BT}{\eta}\left(C_0^2 t + \sum_{j=1}^{K-1} \frac{C_j^2}{r_j} \left(1 - e^{-r_jt} \right) \right).
\end{equation}
Thus, calculating $\cov(x_i, x_j)$ and $\inner{X(t)^2}$ requires that one find the roots of $q'(y)$.  Since $q(y)$ has its unique mode on $(\a_j, \a_{j+1})$ at $y = r_j$, the roots of $q'(y)$ can be efficiently computed using the golden section search algorithm.


\section{Marginalization of Regression Parameters in Gaussian Models}\label{sec:marginf}

The the location-scale form~\eqref{eq:lsmod} of the fBM and GLE models~\eqref{eq:fbm} and~\eqref{eq:gle0} both have likelihoods which can be expressed as linear regressions with multiple responses:
\begin{equation}\label{modelmulti}
\Y_{n\times q} \| \b_{p\times q}, \Sigma_{q\times q}, \t \sim \N_{n\times q}\Big(\X(\t) \b, \V(\t), \Sigma \Big),
\end{equation}
where $\N_{n\times q}$ denotes the matrix normal distribution.  That is, consider the 2-dimensional position process
\begin{equation}\label{eq:lsmod2}
\X(t) = \mmu t + \SSigma^{1/2} \Z(t),
\end{equation}
with $\Z(t) = \big(\Z_1(t), \Z_2(t)\big)$ being iid copies of a 1-dimensional fBM or GLE with parameters $\ap_\tx{fBM} = H$ or $\ap_\tx{GLE} = (\asp, \tau)$.  For increments $\x = (\rv \x N)$, with $\x_n = \X(n \d t) - \X((n-1)\d t)$, the likelihood $\ell(\mmu, \SSigma, \asp \| \x)$ for model~\eqref{eq:lsmod2} is identical to that of~\eqref{modelmulti} with
\begin{align*}
\Y_{N\times 2} & = \x, & \X_{N\times 1}(\t) & = (\d t, \ldots, \d t)', & \b_{1\times 2} & = \mmu, & \t & = \ap, & \V(\t) & = \V_{\ap}.
\end{align*}
The formulas below employ the new parametrization for a greater level of generality.  For instance, the methodology can also be used to replace the linear drift in~\eqref{eq:lsmod2}, $\E{X_i(t)} = \mu_i t$, by a non-linear drift of the form
\[
\E{\X_i(t)} = \sum_{j=1}^p \beta_{ij} f_{j}(t, \t),
\]
for given functions $f_1(t, \t), \ldots, f_p(t, \t)$.  Note that~\eqref{modelmulti} can be rewritten as a multivariate normal using vectorization and the Kronecker product, such that
\[
\vec(\Y) \| \b, \SSigma, \t \sim \N_{nq} \Big( \vec(\X \b), \bm \Sigma \otimes \V \Big).
\]

\subsection{Profile Likelihood}

For every value of $\t$, let
\[
[\Y \| \X]' \V^{-1}[\Y \| \X] = \R = \begin{pmatrix} {\bm s}_{q\times q} & \U' \\ \U_{p\times q} & \T_{p\times p}\end{pmatrix},
\]
and
\begin{align*}
\hat \b_{p\times q} & = (\X' \V^{-1} \X)^{-1} \X' \V^{-1} \Y  & {\bm S}_{q\times q} & = (\Y - \X \hat \b)' \V^{-1} (\Y - \X \hat \b) \\
& = \inv {\T} \U,  & & = {\bm s} - \U' \hat \b.
\end{align*}
The log-likelihood for~\eqref{modelmulti} is
\begin{equation}\label{multill}
\ell(\b, \Sigma, \t \| \Y) = -\frac 1 2 \left[\tr\left\{\SSigma^{-1}(\Y - \X\b)'\V^{-1}(\Y-\X\b)\right\} + n \log(\abs \SSigma) + q \log(\abs \V) \right].
\end{equation}
For fixed $\t$,~\eqref{multill} is maximized at $\hat \b(\t) = \hat \b$ and $\hat \SSigma(\t) = \tfrac 1 n {\bm S}$, where the dependence on $\t$ is through $\X$ and $\V$.  The resulting profile log-likelihood for $\t$ is
\[
\ell_\tx{prof}(\t \| \Y) = -\tfrac 1 2 \Big(nq + n \log(\tfrac 1 n {\bm S}) + q\log(\abs{\V}) \Big).
\]

\subsection{Conjugate Prior}

For the linear regression model~\eqref{modelmulti}, the conjugate prior is
\begin{equation}\label{eqconjprior}
\begin{split}
\t & \sim \pi(\t) \\
\SSigma \| \t & \sim \tx{Inv-}\mathcal W(\PPsi, \nu) \\
\b \| \SSigma, \t & \sim \N_{p\times q}(\LL, \OO^{-1}, \SSigma),
\end{split}
\end{equation}
where
\[
p(\SSigma \| \t) = \Xi(\PPsi, \nu) \abs{\SSigma}^{-(\nu+q+1)/ 2} \exp\{-\tfrac 1 2 \tr(\PPsi \SSigma^{-1})\}
\]
and
\[
\Xi(\PPsi,\nu) = \frac{\abs{\PPsi}^{\nu/2}}{\sqrt{2^{\nu q}} \Gamma_q(\tfrac \nu 2)}, \quad \Gamma_q(a) = \pi^{q(q-1)/4} \prod_{j=1}^q \Gamma[a + \tfrac 1 2(1-j)].
\]
Note that $\PPsi = \PPsi(\t)$, $\nu = \nu(\t)$, $\LL = \LL(\t)$, and $\OO = \OO(\t)$ can all be made to depend on $\t$.  Conditioned on $\t$, $(\b,\SSigma)$ in~\eqref{eqconjprior} have a Matrix-Normal-Inverse-Wishart (MNIW) distribution:
\[
\b, \SSigma \| \t \sim \tx{MNIW}(\LL, \OO^{-1}, \PPsi, \nu).
\]

The conjugate posterior distribution of all parameters is
\begin{equation}\label{eqconjpost}
\begin{split}
\t \| \Y & \sim p(\t \| \Y) \propto \pi(\t) \frac{\Xi(\PPsi, \nu)}{\Xi(\hat\PPsi,\hat\nu)} \left(\frac{\abs{\OO}(2\pi)^{-n}}{\abs{\hat \OO}\abs{\V}}\right)^{q/2}\\
\b, \SSigma \| \t, \Y & \sim \tx{MNIW}(\hat \LL, \hat \OO^{-1}, \hat \PPsi, \hat \nu),
\end{split}
\end{equation}
where
\begin{align*}
\hat \OO & = \OO + \T, & \hat \LL & = \hat \OO^{-1}(\T \hat \b+ \OO \LL), \\
\hat \nu & = \nu + n, & \hat \PPsi & = \PPsi + {\bm S} + \hat \b' \T \hat \b + \LL' \OO \LL - \hat \LL' \hat \OO \hat \LL.
\end{align*}
In this paper, we have used the improper location-scale invariant prior
\[
\pi(\b, \SSigma \| \t) \propto \abs{\SSigma}^{-(\nu + q + 1)/2}.
\]
The posterior~\eqref{eqconjpost} follows upon setting $\LL = \bm 0$, $\OO = \bm 0$, $\PPsi = \bm 0$, and $\hat \nu = \nu + n-p$.


\subsection{Marginal Likelihoods}

If the conjugate prior distribution~\eqref{eqconjprior} is proper, the data has marginal likelihood
\[
f(\bm Y) = \int p(\bm Y \| \bm \Theta) \pi(\bm \Theta) \ud \bm \Theta.
\]
This quantity is required to calculate the posterior model probabilities in Section~\ref{sec:modelcomp}.  To compute it efficiently for a linear regression model, note that
\[
\log f(\bm Y) = K - \frac{nq} 2 \log(2\pi),
\]
where
\[
\exp(K) = \int \pi(\t) \frac{\Xi(\PPsi, \nu)}{\Xi(\hat\PPsi,\hat\nu)} \left(\frac{\abs{\OO}(2\pi)^{-n}}{\abs{\hat \OO}\abs{\V}}\right)^{q/2} \ud \t.
\]
Thus, $f(\Y)$ is closely related to the normalizing constant for $p(\t \| \Y)$ in~eqref{eqconjpost}.  It can be estimated with a Monte Carlo sample from $p(\t \| \Y)$~\citep[e.g.][]{marin-robert10}, or in the case of fBM and GLE, with deterministing integrals over $\t_\tx{fBM} = H$ and $\t_\tx{GLE} = (\asp, \tau)$ respectively.


\subsection{Method-of-Moments Estimator for the MNIW Distribution}\label{sec:mniwmom}

In Section~\ref{sec:hmapprox}, we approximated the prior of interest by a conjugate prior of the form~\eqref{eqconjprior} to greatly accelerate calculations.  The following method-of-moments estimate of the parameters of an MNIW distribution is a major component of this approximation.  The remaining details are provided in Subsection~\ref{sec:conjapprox}.

Let $(\b, \SSigma) \sim \tx{MNIW}(\LL, \UU, \PPsi, \nu)$.  Then for $\eta = \nu-q-1$ we have
\begin{align*}
E[\b] & = \LL & \var\big(\vec(\b)\big) & = \frac{\PPsi \otimes \UU}{\eta} = E[\SSigma] \otimes \UU \\
E[\SSigma] & = \frac{\PPsi}{\eta} & \cov(\SSigma_{ij}, \SSigma_{kl}) & = \frac{2\PPsi_{ij}\PPsi_{kl} + \eta(\PPsi_{ik}\PPsi_{jl} + \PPsi_{il}\PPsi_{kj})}{\eta^2(\eta+1)(\eta-2)} \\
\cov(\b, \SSigma) & = \bm 0 & \var(\SSigma_{ii}) & = \frac{2\PPsi_{ii}^2}{\eta^2(\eta-2)} = \frac{2\big(E[\SSigma_{ii}]\big)^2}{\eta-2}.
\end{align*}
This suggests the following method-of-moments estimator for the parameters of the MNIW distribution.  Let $\bar \b_{p\times q}$ and $\bar \SSigma_{q\times q}$ be estimates of $E[\b]$ and $E[\SSigma]$.  Let ${\bm S}_i$ be an estimate of $\var(\b_i)_{p\times p}$, where $\b_i$ is the $i$th column of $\b$, and $d_j$ be an estimate of $\var(\SSigma_{jj})$.  Then the MNIW parameter estimates are:
\begin{align*}
\hat \LL & = \bar \b & \hat \nu & = q + 3 + \frac 2 q \sum_{j=1}^q \bar \SSigma_{jj}^2/d_j \\
\hat \PPsi & = (\hat \nu - q - 1)\hat \SSigma & \hat \UU & = \frac 1 p \sum_{i=1}^p {\bm S}_i/\bar \SSigma_{ii}.
\end{align*}


\subsection{Conjugate Approximation to the Hierarchical Prior}\label{sec:conjapprox}

In Section~\ref{sec:hmapprox}, we wish to approximate the distribution $\pi_{\tx{test}}(\t)$ by a conjugate prior of the form~\eqref{eqconjprior}.  The conjugate prior is defined on the parametrization $\t = (\ap, \mmu, \SSigma)$.  However, the MCMC sampler described in Section~\ref{sec:hmapprox} produces draws of 
$\tilde \t = (\ap, \mmu, \lsr)$, where
\begin{align*}
\ap_\tx{fBM} & = H & \mmu & = (\mu_1, \mu_2) \\
\ap_\tx{GLE} & = \big(\log(\tau)/\g, \log(\tau)\big) & \lsr & = \big(\log(\s_1), \log(\s_2), \rho\big).
\end{align*}
We found that $\pi_\tx{test}(\tilde \t)$ is well approximated by a multivariate normal:
	\begin{equation}\label{eqpriornormapprox}
	\begin{bmatrix} \ap \\ \mmu \\ \lsr \end{bmatrix} \approx \N\left\{\begin{bmatrix} \bm m_\ap \\ \bm m_\mmu \\ \bm m_{\lsr} \end{bmatrix}, \begin{bmatrix} \bm V_{\ap\ap} & \bm V_{\ap\mmu} & \bm V_{\ap\lsr} \\ \bm V_{\mmu\ap} & \bm V_{\mmu\mmu} & \bm V_{\mmu\lsr} \\ \bm V_{\lsr\ap} & \bm V_{\lsr\mmu} & \bm V_{\lsr\lsr}\end{bmatrix} \right\}.
	\end{equation}
On the other hand, the conjugate prior specifies the conditional distribution
\[
\mmu, \SSigma \| \ap \sim \tx{MNIW}(\LL_\ap, \UU_\ap, \PPsi_\ap, \nu_\ap),
\]
where $\SSigma = \left[\begin{smallmatrix} \s_1^2 & \s_1\s_2\rho \\ \s_1\s_2\rho & \s_2^2 \end{smallmatrix}\right]$.  In~\ref{sec:mniwmom}, we provided a method-of-moments estimate for $\LL_\ap$, $\UU_\ap$, $\PPsi_\ap$, and $\nu_\ap$ based on $E[\mmu \| \ap]$, $\var(\mu_i \| \ap)$, $E[\SSigma \| \ap]$, and $\var(\SSigma_{ii} \| \ap)$, for $i = 1,2$.  Using results for multivariate normals, the conditional moments of $\mmu$ are approximated by
\begin{equation}\label{eqcondmom}
\begin{split}
E[\mmu \| \ap] & = \bm m_{\mmu \| \ap} \approx \bm m_\mmu + \bm V_{\mmu\ap}V_{\ap\ap}^{-1}(\ap - \bm m_\ap) \\
\var(\mmu \| \ap) & = \bm V_{\mmu \| \ap} \approx \bm V_{\mmu\mmu} - \bm V_{\mmu\ap}\bm V_{\ap\ap}^{-1}V_{\ap\mmu}.
\end{split}
\end{equation}
As for the conditional moments of $\SSigma$, we note that for $\SSigma = [\Sigma_{ij}]$ and $\lsr = (\varpi_1, \varpi_2, \varpi_3)$, we have
\begin{align*}
\Sigma_{11} & = \exp(2\varpi_1) & \Sigma_{22} & = \exp(2\varpi_2) & \Sigma_{12} & = \varpi_3\exp(-\varpi_1-\varpi_2).
\end{align*}
To calculate the relevant moments of these random variables, we use~\eqref{eqpriornormapprox} to approximate the conditional distribution of $\lsr$ by
\[
\lsr \| \ap \approx \N(\bm m, \bm V),
\]
with $\bm m = \bm m_{\lsr \| \ap}$ and $\bm V = \bm V_{\lsr \| \ap}$ derived analogously to~\eqref{eqcondmom}.  Thus, the conditional distribution of $\Sigma_{ii}$ is approximately log-normal, such that
\begin{align*}
E[\Sigma_{ii} \| \ap] & = m_{\Sigma_{ii} \| \ap} \approx \exp(2 m_{i} + 2 V_{ii}) \\
\var(\Sigma_{ii} \| \ap) & = V_{\Sigma_{ii}\|\ap} \approx [\exp(4V_{ii}) - 1] \exp(4 m_i + 4 V_{ii}).
\end{align*}
As for $E[\Sigma_{12} \| \ap] = E[\Sigma_{21} \| \ap]$, assuming that $\lsr$ is conditionally normal we have
\begin{align*}
E[\Sigma_{12} \| \ap] & \approx \int \frac{\varpi_3 \exp(-\varpi_1 -\varpi_2)}{(2\pi)^{d/2}\abs{\bm V}} \exp\{-\tfrac 1 2 (\lsr - \bm m)'{\bm V}^{-1}(\lsr - \bm m)\} \ud \lsr \\
& = C \int \frac{\varpi_3}{(2\pi)^{d/2}\abs{\bm V}} \exp\{-\tfrac 1 2 (\lsr - \tilde{\bm m})'{\bm V}^{-1}(\lsr - \tilde{\bm m})\} \ud \lsr,
\end{align*}
where, for $\bm a' = [ -1, -1, 0]$, we have $\tilde {\bm m} = \bm m + \bm V \bm a$ and $C = \exp\{\bm a' \bm m + \tfrac 1 2 \bm a' \bm V \bm a\}$.  Thus,
\begin{align*}
E[\Sigma_{12} \| \ap] & = m_{\Sigma_{12} \| \ap} \approx C \tilde m_3 \\
& = (m_3 - V_{31} - V_{32})\times\exp\{\tfrac 1 2 (V_{11} + V_{22} + V_{12}) - m_1 - m_2\} .
\end{align*}
Thus, our conjugate prior approximation to distribution of $\pi_\tx{test}(\t)$ obtained from the MCMC sampling scheme in Section~\ref{sec:hmapprox} is
\begin{align*}
\ap & \sim \N(\bm m_\ap, \bm V_{\ap\ap}) \\
\mmu, \SSigma \| \ap & \sim \tx{MNIW}(\LL_\ap, \UU_\ap, \PPsi_\ap, \nu_\ap),
\end{align*}
where $\LL_\ap$, $\UU_\ap$, $\PPsi_\ap$, and $\nu_\ap$ are derived from $m_{\Sigma_{ij} \| \ap}$ and $V_{\Sigma_{ii} \| \ap}$ as in~\ref{sec:mniwmom}.


\section{Sources of Instrumental Measurement Error}\label{sec:error}

\cite{savin-doyle05} discuss a number of sources of experimental error in particle-tracking microrheology.  Those of primary concern for our experimental setup are: (1) additional noise in the observed data due to instrument vibration, and (2) diminished spatial resolution for particles located at the edge of the microscope's focal plane, which results in coarse discretization of the measurements of particle position.

For (2), such paths can be easily identified by the experimentalist, and thus all paths on the edge of the focal plane have been removed before proceeding with the 76 paths in our analysis.  To quantify the impact of (1), the standard experimental procedure is to allow the tracer bead particles to dry onto their glass mountings, immobilizing them completely (no displacement).  Thus, the observed displacement of these immobilized beads is purely due to instrumental error. 

\begin{figure}[htb!]
	\centering
		\includegraphics[width=1.00\textwidth]{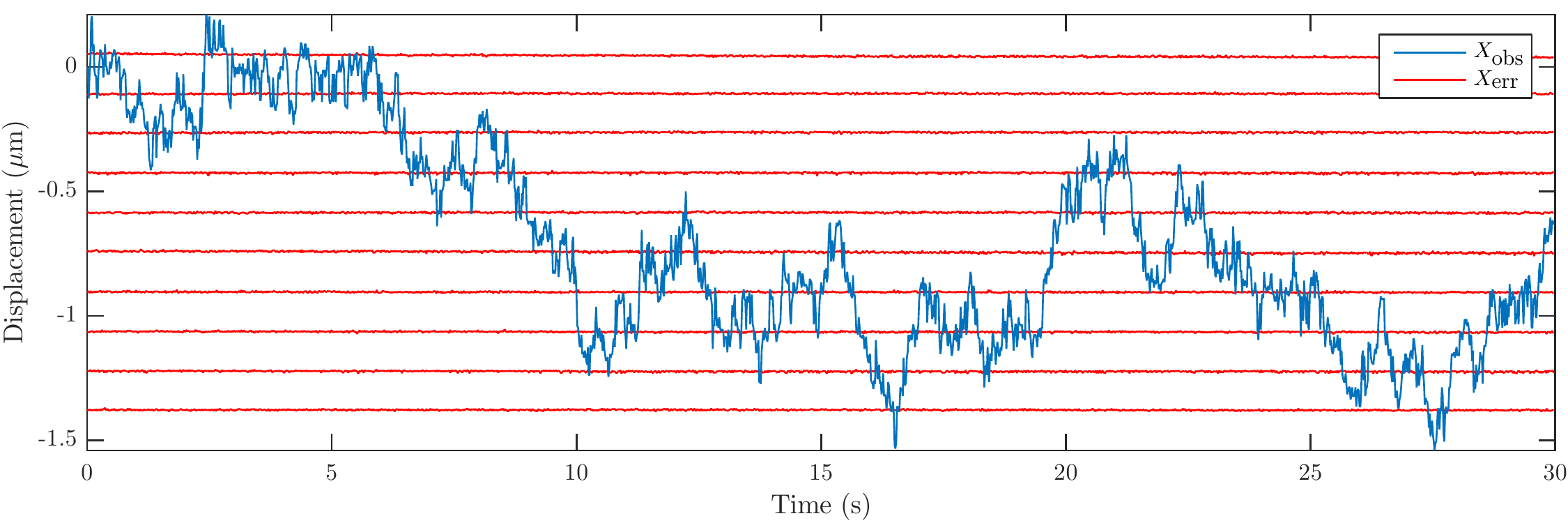}
	\caption{Comparison of 10 one-dimensional displacement curves of pure measurement error ($X_\tx{err}$) to a single particle trajectory comprised of both signal and noise ($X_\tx{obs}$).}
	\label{fig:err}
\end{figure}

Figure~\ref{fig:err} displays a number of such ``pure error'' trajectories along a single spatial dimension, along with a typical trajectory for an unstuck bead, which consists of both signal and noise.
For the current experimental setup, Figure~\ref{fig:err} suggests that the measurement error is negligible.  Indeed, we have verified with simulated trajectories from fBM and GLE models that adding measurement error on the scale of Figure~\ref{fig:err} does not alter parameter estimates.  Therefore, we conducted the analysis in this article under the assumption that the measurement error in our experiments was effectively zero.


\section{Sensitivity of Bayesian Model Selection to the Choice of Prior}\label{sec:badprior}

To illustrate this problem in our particular setting, we picked one of the datasets in Figure~\ref{fig:data} (DS-1), denoted $\x_\obs$, and fit the MLE for fBM and GLE-200 models.  These estimates were taken as true parameter values, $\t_\tx{fBM} = (\mmu, \SSigma, H)$ and $\t_\tx{GLE} = (\mmu, \SSigma, \ap, \tau)$, from which we simulated 10 datasets from each of the two models.

Next, for each of the 20 simulated datasets $\x_\tx{sim}^{(j)}$, $j = 1, \ldots, 20$, and each model $M_i$, $i \in \{\tx{fBM}, \tx{GLE}\}$, we computed posterior model probabilities $p(M_i \| \x_\tx{sim}^{(j)})$ under equal prior odds $\pi(M_i) = \tfrac 1 2$, and proper parameter priors $\pi(\t_i \| M_i)$ at two opposing extremes:
\begin{enumerate}
\item A noninformative but proper prior $\pi_{\tx{NI}}(\t_i \| M_i)$.

\noindent This prior was chosen to closely resemble the improper priors~\eqref{eq:specprior} used for the independent trajectory analysis in Section~\ref{sec:modelfit}.  These proper posteriors were of the conjugate form~\eqref{eq:conjprior}, with $\bm\Psi = \left[\begin{smallmatrix} 10^{-20} & 0 \\ 0 & 10^{-20} \end{smallmatrix}\right]$, $\nu = 10^{-20}$, $\LL = (0, 0)$, and $\kappa = 1000$.
\item An overly-informative prior $\pi_\tx{2x}(\t_i \| M_i)$.

\noindent This prior is implied by the \emph{posterior Bayes factor} of~\cite{aitkin91}.  For each dataset $\x_\tx{sim}^{(j)}$, it is defined as
\[
\pi_\tx{2x}(\t_i \| M_i) = p(\t_i \| \x_\tx{sim}^{(j)}, M_i),
\]
the posterior distribution obtained from the improper priors~\eqref{eq:specprior} of Section~\ref{sec:modelfit}.
The marginal likelihood for $\x_\tx{sim}^{(j)}$ is
\begin{equation}\label{eq:marglik}
f(\x_\tx{sim}^{(j)} \| M_i) = \int p(\x_\tx{sim}^{(j)} \| \t_i, M_i) \pi_\tx{2x}(\t_i \| M_i) \ud \t_i,
\end{equation}
which uses the data twice: once in the conditional model $p(\x \| \t_i, M_i)$, and once in the prior.
\end{enumerate}

\begin{figure}[htb!]
	\centering
		\includegraphics[width=1.00\textwidth]{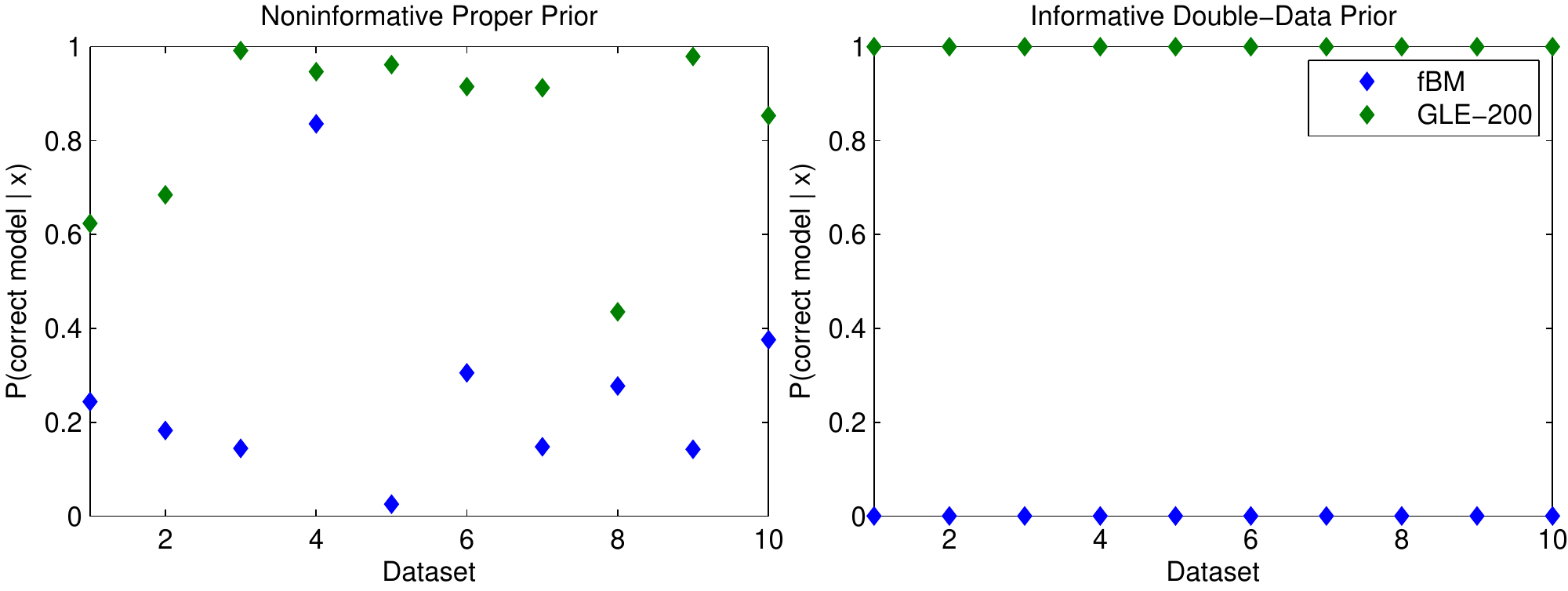}
	\caption{Posterior probability of picking the correct model for simulated fBM and GLE-200 datasets for two poor choices of prior.}
	\label{fig:badprior}
\end{figure}

Figure~\ref{fig:badprior} displays the posterior probabilities $p(M_i \| \x_\tx{sim}^{(j)})$ for each of the 20 simulated datasets. 
Both priors clearly fail to separate the two models, but for different reasons.  The poor performance of the noninformative prior $\pi_\tx{NI}(\t_i \| M_i)$ is a consequence of Lindley's paradox~\citep{lindley57}.  Failure of the overly-informative prior $\pi_\tx{2x}(\t_i \| M_i)$ is due to overfitting.  That is, if the data is not used ``twice'' but ``$m$ times'', as $m \to \infty$ the marginal likelihood~\eqref{eq:marglik} converges to $p(\x_\tx{sim}^{(j)}, \hat \t_i, M_i)$, where $\hat \t_i$ is the MLE of model $M_i$.  Since the log-densities $\log(p(\x_\tx{sim}^{(j)} \| \hat \t_i, M_i))$ are normal, each consists of a sum of squared residuals and a log-determinant.  A close look at the simulated data reveals that the residuals for the fBM and GLE-200 models at their MLEs are virtually the same.  
However, for any fixed number of modes, the GLE model is asymptotically \emph{diffusive}.  Thus, while fBM increments have power law decay, the GLE-200 increments have power law intermediate behavior, but exponential decay.  This translates to more sharply decaying eigenvalues of the corresponding variance matrix, and thus smaller log-determinant penalty.  It is for this reason that the double data prior emphatically favors the GLE-200 model, even for data simulated under fBM.

\end{appendices}


\bibliographystyle{stdref}
\bibliography{microrheology_aoas}

\end{document}